\documentclass{aa}  

\usepackage{graphicx}
\usepackage{txfonts}
\usepackage{color}

\usepackage{natbib}
\bibpunct{(}{)}{;}{a}{}{,} 
\begin{document} 

    \title{Frequency analysis of the first-overtone RR Lyrae stars based on the Extended Aperture Photometry from the K2 data}
   \titlerunning{First-overtone RR Lyrae stars in the K2 data}

   \author{H. Netzel
          \inst{1,2,4}\fnmsep\thanks{henia@netzel.pl}
          \and
          L. Moln\'ar\inst{1,2,3}
          \and
          E. Plachy\inst{1,2,3}
          \and
          J. M. Benk\H{o} \inst{1,2}
          }

   \institute{Konkoly Observatory, Research Centre for Astronomy and Earth Sciences, E\"otv\"os Lor\'and Research Network (ELKH); MTA Centre of Excellence, H-1121 Konkoly Thege Mikl\'os \'ut 15-17, Budapest, Hungary
         \and
             MTA CSFK Lend\"ulet Near-Field Cosmology Research Group, H-1121 Konkoly Thege Mikl\'os \'ut 15-17, Budapest, Hungary
         \and
             ELTE E\"otv\"os Lor\'and University, Institute of Physics, 1117, P\'azm\'any P\'eter s\'et\'any 1/A, Budapest, Hungary
    \and
    ELTE E\"otv\"os Lor\'and University, Gothard Astrophysical Observatory, Szent Imre h. u. 112, 9700, Szombathely, Hungary
             }

   \date{Received September 15, 1996; accepted March 16, 1997}

 
  \abstract
   {Additional low-amplitude signals are observed in many RR Lyrae stars, beside the pulsations in radial modes. The most common ones are short-period signals forming a period ratio of around 0.60--0.65 with the first overtone, or long-period signals forming a period ratio of around 0.68. The RR Lyrae stars may also exhibit quasi-periodic modulation of the light curves, known as the Blazhko effect.}
   {We used the extensive sample of the first-overtone RR Lyrae stars observed by the {\it Kepler} telescope during the K2 mission to search for and characterize these low-amplitude additional signals. K2 data provides space-based photometry for a statistically significant sample. Hence this data is excellent to study in detail pulsation properties of RR Lyrae stars.}
   {We used K2 space-based photometry for RR Lyrae candidates from Campaigns 0--19. We selected RR Lyrae stars pulsating in the first overtone and performed a frequency analysis for each star to characterize their frequency contents.} 
   {We classified 452 stars as first-overtone RR Lyrae. From that sample, we selected 281 RR$_{0.61}$ stars, 67 RR$_{0.68}$ stars, and 68 Blazhko stars. We found particularly interesting stars which show all of the above phenomena simultaneously. We detected signals in RR$_{0.61}$ stars that form period ratios lower than observed for the majority of stars. These signals likely form a new sequence in the Petersen diagram, around a period ratio of 0.60. In 32 stars we detected additional signals that form a period ratio close to that expected in RRd stars, but the classification of these stars as RRd is uncertain. We also report a discovery of additional signals in eight stars that form a new group in the Petersen diagram around the period ratio of 0.465--0.490. The nature of this periodicity remains unknown.}
   {}

   \keywords{Stars: variables: RR Lyrae --
                Stars: oscillations (including pulsations) --
                Stars: horizontal-branch
               }

   \maketitle
%

\section{Introduction}
RR Lyrae stars are low-mass population II pulsating stars from the classical instability strip. The majority of them pulsate in the radial fundamental mode (RRab), first overtone (RRc), or the fundamental and first-overtone modes simultaneously (RRd). Double-mode pulsations in the fundamental and second-overtone modes are also known, but such stars are relatively rare \citep{benko2010,moskalik2013}. Similarly, triple-mode pulsations in radial modes is a rare phenomenon reported in a few stars \citep{molnar2012,jurcsik2015,soszynski2014}. For many years the only unexplained phenomenon occurring in RR Lyrae stars was the presence of the Blazhko effect. 

The Blazhko effect is a quasi-periodic modulation of the amplitude and/or phase of the pulsations \citep{blazhko}. It is more common among RRab than in RRc stars. \cite{jurcsik2009} inferred an incidence rate of nearly 50 per cent of the Blazhko effect in RRab stars. The incidence rate is lower for RRc stars. In the Galactic bulge, it is estimated to be below 6 per cent based on the ground-based data \citep{netzel_blazhko}. The Blazhko effect is also known among RRd stars \citep{jurcsik2015,smolec2015,plachy2017rrdbl,carrell2021}. The origin of the Blazhko effect remained a mystery for over a century and many mechanisms have been proposed, but none fully explains the properties. Only recently, \cite{kollath2021} was able to produce amplitude modulation similar to the Blazhko effect in the theoretical model of an RRab star via interactions between the fundamental mode and the 9th overtone, although it does not explain all of the aspects of the Blazhko effect observed in RRab stars. Still, the explanation of the Blazhko effect in RRc stars is missing.

Thanks to new photometric observations from ground-based large-scale surveys, like the Optical Gravitational Lensing Experiment \citep[OGLE,][]{ogleiv} and space-based missions like CoRoT, {\it Kepler} and TESS, we gained new insights into the pulsations of RR Lyrae stars. A variety of low-amplitude additional signals that do not correspond to radial modes were detected. The most numerous group of RR Lyrae stars with additional periodicities consists of RRc and RRd stars with low-amplitude short-period signals forming a period ratio of around 0.60--0.64 with the first overtone, the so-called RR$_{0.61}$ stars. The first member of this group has been an RRd star, AQ Leo, observed with the MOST space telescope \citep{gruberbauer2007}. Currently, over a thousand RR$_{0.61}$ stars are known. They were observed both in ground-based  \citep[e.g.][]{jurcsik2015,smolec2017,netzel_census} and space-based surveys  \citep[e.g.][]{moskalik2015,molnar2015,molnar2022}. The incidence rate of RR$_{0.61}$ stars increases with the increasing quality of the photometry. 
The incidence rate is below 60 per cent for ground-based data, but exceeds that for space-based photometry, suggesting that this multi-mode type of pulsation is a common feature among first-overtone RR Lyrae stars. Interestingly, this additional signal was not detected in RRab stars, where the first overtone is not present. An explanation for the nature of RR$_{0.61}$ stars was proposed by \cite{dziembowski2016}, who suggested that the observed signals are due to harmonics of non-radial modes of degrees $\ell = 8$ and 9. We note that similar signals have been observed in the first-overtone classical Cepheids \citep[see e.g.][and references therein]{soszynski2010_rrl,rajeev}. The model proposed by \cite{dziembowski2016} explains these analogous signals in classical Cepheids as harmonics of non-radial modes of degrees 7, 8, and 9. Interestingly, \cite{plachy2021} reported the detection of an analogous signal in an overtone anomalous Cepheid.

Another group of multi-mode RR Lyrae stars are the so-called RR$_{0.68}$ stars \citep{netzel068}. In these stars, the dominant pulsation mode is the first overtone. The additional signal has a longer period and forms a period ratio with the first overtone of about 0.686, meaning that its period is longer than the period corresponding to the fundamental mode. Analogous signals were also reported in classical Cepheids by \cite{suveges.anderson2018}. The nature of this additional signal still lacks an explanation.

Additional low-amplitude signals were detected in RRab stars as well. \cite{smolec.prudil2016} reported the discovery of a group of long-period RRab stars with additional shorter-period signals. They suggested that these stars can be explained as pulsating in fundamental mode and first overtone. \cite{prudil2017} found a group of RR Lyrae stars that have an additional signal of a shorter period and are located in the Petersen diagram at the short-period extension of an RRd sequence. The origin of the observed signals remains unknown.

Observations carried out with the {\it Kepler} satellite resulted in many discoveries connected to all these phenomena in RR Lyrae stars. Continuous sampling led to the discovery of period doubling in RRab stars showing the Blazhko effect \citep{kolenberg2010,szabo2010} or to the detection of additional signals in RR Lyrae stars \citep{benko2010,moskalik2015}. The original {\it Kepler} field could no longer be monitored after two reaction wheels failed. A new mission named K2 followed, monitoring subsequent fields along the ecliptic plane \citep{howell2014}. Observations of these fields, known as Campaigns, lasted around 80 days for each field, with an exception of Campaign 19 which lasted around 15 days. The challenges in data processing came with the new mode of observation. \cite{plachy2019} suggested an analysis method optimized for detecting RR Lyrae stars. The method called Extended Aperture Photometry was then automatized and applied to Campaigns 0--19 by \cite{bodi2021}.

In this work, we aim to characterize the phenomena that occur in first-overtone RR Lyrae stars, such as the Blazhko modulation and the presence of additional signals using the K2 light curves extracted by \cite{bodi2021}.

The paper is structured as follows. In Sec.~\ref{sec.methods} we describe the method. We present results in Sec.~\ref{Sec.results} and discuss them in Sec~\ref{Sec.discussion}. Conclusions are collected in Sec.~\ref{Sec.conclusions}.

\begin{figure*}
    \centering
    \includegraphics[width=\textwidth]{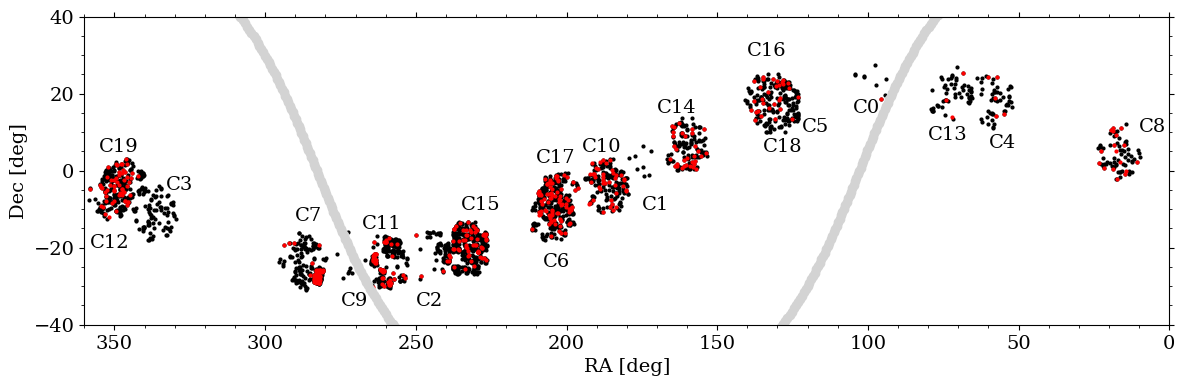}
    \caption{Positions of all RR Lyrae candidates are plotted with black symbols. Stars classified as RRc are plotted with red symbols.}
    \label{fig:all_coords}
\end{figure*}

\section{Methods}\label{sec.methods}
We used light curves of RR Lyrae candidates observed in Campaigns 0 to 19 of the K2 mission, as prepared by \cite{bodi2021}. First, we performed a classification based on light curve shapes to select RRc stars from the whole sample. Second, we carried out the frequency analysis of the classified RRc stars. We searched for stars showing the Blazhko effect and for those that feature additional signals.

\subsection{Data and classification}
The sample of light curves prepared by \cite{bodi2021} consists of 3917 light curves for K2 Campaigns 0 to 19. Some of the stars were observed during multiple campaigns; hence the number of unique stars in the whole sample is 3057. The positions of these stars in sky coordinates are presented in Fig.~\ref{fig:all_coords} with black points.

Classification of the 3057 candidate RR Lyrae stars into subtypes RRc and RRab was carried out in two stages. In the first step, we used Fourier transforms to find the dominant periodicity for each object. Then we fitted a Fourier series to each light curve in the form:

\begin{equation}
    \label{eq.series}
    F(t)=A_0 + \sum_k A_k \sin(2\pi f_k t + \phi_k),
\end{equation}
where $A_0$ is a mean brightness, $A_k$, $f_k$, and $\phi_k$ are the amplitude, frequency, and phase of the $k$-th harmonic of the highest signal in the power spectrum. Only harmonics fulfilling the criterion $A_k/\sigma_k>4.0$ were fitted to the data. From the fit, we obtained Fourier coefficient, $R_{21}$, defined as amplitude ratio \citep{simon.lee1981}:
\begin{equation}
    \label{eq.ak1}
    R_{21}=\frac{A_2}{A_1},
\end{equation}

\begin{figure}
    \centering
    \includegraphics[width=\columnwidth]{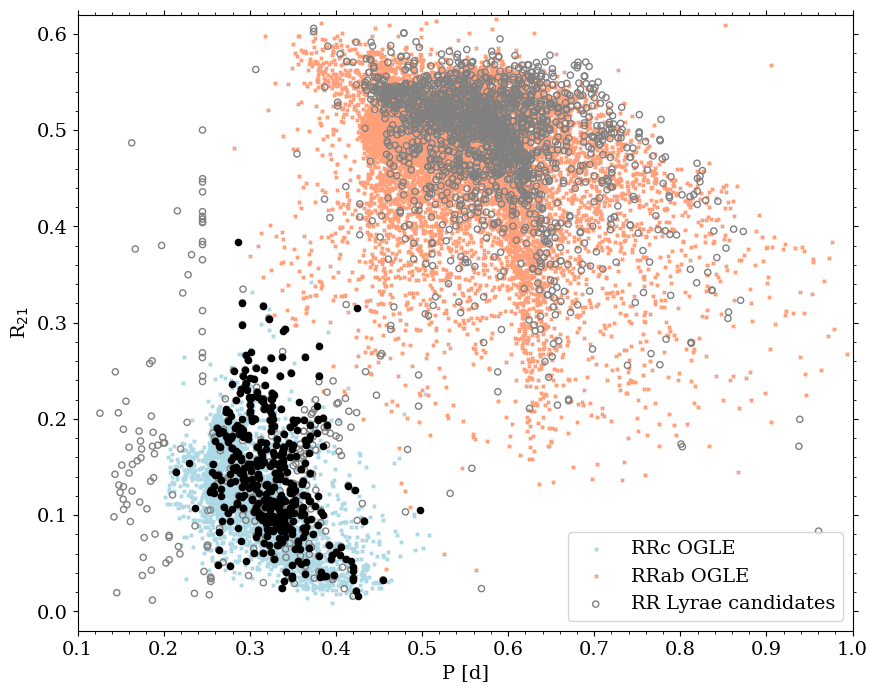}
    \caption{Fourier coefficients for RR Lyrae candidates are plotted with black points. We plotted RRab and RRc stars observed by the OGLE project towards the Galactic bulge with light blue and orange points \protect\citep{soszynski2014}.}
    \label{fig:fdp}
\end{figure}

Fourier coefficient, $R_{21}$, is presented in Fig.~\ref{fig:fdp} with grey open circles. For reference, we included RRab and RRc stars observed by the OGLE project towards the Galactic bulge \citep{soszynski2014}. Based on Fourier coefficient $R_{21}$, we performed the initial classification of stars for RRc and RRab subtypes. Several stars with period or $R_{21}$ values atypical for RR Lyrae stars are not presented in Fig.~\ref{fig:fdp} and were not analyzed further. 

We visually inspected light curves phased with the dominant periodicity in the second classification step. This step allowed us to exclude from the RRc sample objects that are not RRc stars but have similar Fourier coefficient values, e.g., eclipsing binaries. In particular, we excluded objects in which the only periodicity originates from the instrumental signal connected to the K2 observations (see points forming a line of the constant period around 0.25\,d in Fig.~\ref{fig:fdp}).

We classified 452 stars as RRc. These stars are plotted with red symbols in Fig.~\ref{fig:all_coords} and with black symbols in Fig.~\ref{fig:fdp}. Further analysis focuses only on those stars. We note, that for stars observed during multiple Campaigns, we did not merge the data from individual Campaigns. Typically, these Campaigns were far apart from each other. Due to technical reasons, we did not join such datasets. For the further analysis, for each star we chose data from only one Campaign, in which the instrumental trends were the least significant.

\subsection{Frequency analysis}
To do the frequency analysis, we followed the standard consecutive prewhitening method. After prewhitening with the dominant signal and its harmonics, we searched for additional signals in the Fourier spectrum of the residual light curve. If an additional signal was detected, it was added to the Fourier series in the form of Eq.~\ref{eq.series} either as an independent frequency or as a linear combination of frequencies already included in the fit. Then, we subtracted the Fourier fit from the data and repeated the inspection of the residuals. The Nyquist frequency for the analyzed data is around 24 c/d. We inspected the frequency range of 0--20 c/d. Again, only terms fulfilling the criterion $A/\sigma_A > 4$ remained in the final light curve solution.

In some stars, we observe significant long-term trends in the data of likely instrumental origin. This manifests in the frequency spectra as signals at low frequencies. Such signals increase the noise level and hamper the search for signals at low amplitudes. We removed such trends by fitting a spline function to the residuals and subtracting it from the data. 

RRc stars often show irregular period changes, manifesting as additional signals close to the first-overtone frequency \citep{jurcsik2001}. Often these signals are unresolved with the first-overtone frequency. We considered two signals to be resolved within the resolution of the Fourier transform when their separation was larger than $2/\Delta T$, where $\Delta T$ is the time baseline. We included only resolved signals in the solution.

The three most common groups of RRc stars with additional signals are RR$_{0.61}$, RR$_{0.68}$ and Blazhko stars. In RR$_{0.61}$ stars we observe additional signals of period, $P_{0.61}$, shorter than the first-overtone period, $P_{\rm 1O}$, and typically $P_{0.61} \in \left\langle 0.60, 0.64 \right\rangle P_{\rm 1O}$. We allowed for the larger margin and classified stars as RR$_{0.61}$ when $P_{0.61} \in \left\langle 0.57, 0.65 \right\rangle P_{\rm 1O}$. Often many signals fulfilled this criterion in a single star (see Sec.~\ref{Sec.061}). We included all resolved signals in the fit.

A common feature of RR$_{0.61}$ stars is the presence of subharmonics of the additional signals, $f_{0.61}$, at around $0.5f_{0.61}$ and $1.5f_{0.61}$. The subharmonics often show complex structures in the frequency spectra (see Sec.~\ref{Sec.061}). The structures of signals at $f_{0.61}$ and its subharmonics often consist of many unresolved peaks, which makes the complete prewhitening, to the point where only the noise is left in the frequency spectrum of residuals, impossible.  

The additional signal in the RR$_{0.68}$ stars has a period longer than the first-overtone period. The period ratio between the two periodicities is around 0.686, with a relatively small scatter around this value observed among currently known RR$_{0.68}$ stars. Since we do not know the range of possible deviation from $P_{0.68}/P_{\rm 1O} \approx 0.686$, we again allowed a larger margin of period ratios. We classified stars as RR$_{0.68}$ when the period ratio between the first-overtone period and the period of the additional signal was in a range of 0.66--0.71.

The Blazhko effect can normally be detected from visual inspection of the light curve (see examples in Sec.~\ref{Sec.blazhko}). In the frequency spectrum, it manifests as modulation multiplets centred at the pulsation frequency and its harmonics. The frequency separation between the sidepeaks and the pulsation frequency corresponds to the frequency of modulation, $f_B$, and its period, $P_B$. The structure and amplitudes of sidepeaks depend on the modulation properties \citep{benko2011}.

\begin{figure}
    \centering
    \includegraphics[width=\columnwidth]{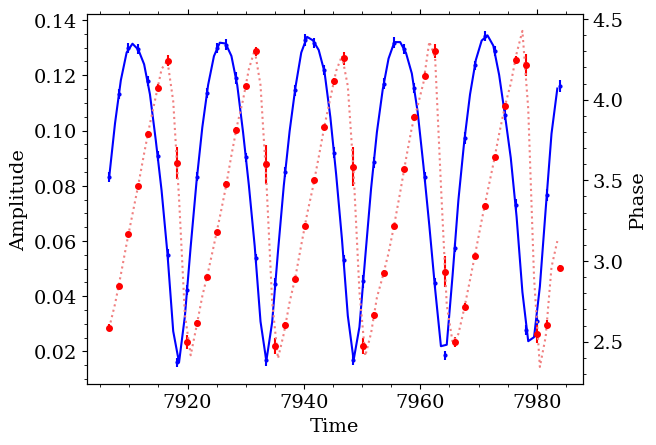}
    \includegraphics[width=\columnwidth]{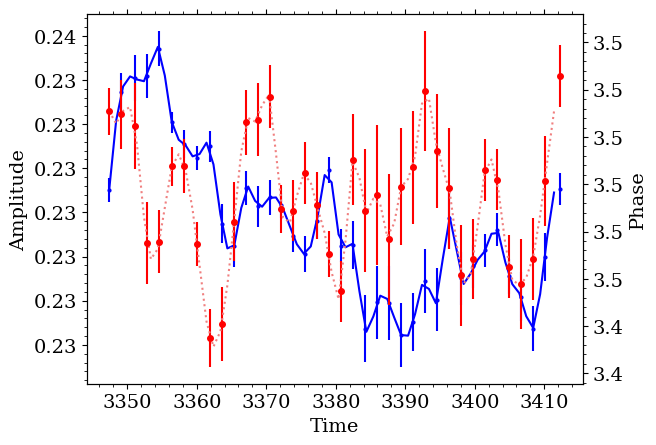}
    \caption{Amplitude (blue) and phase (red) changes over time in two stars with detection of signals close to the first-overtone frequency. The star presented in the top panel, EPIC\,248553635, was classified as Blazhko, whereas the star presented in the bottom panel, EPIC\,212595129, was not.}
    \label{fig:mod}
\end{figure}

We classified candidates for Blazhko stars based on the visible light curve modulation or based on the presence of sidepeaks in frequency spectra. We note that in some stars the modulation is visible in the light curve, but the data coverage is too short to detect resolved sidepeaks in the frequency spectra. However, simply the presence of sidepeaks was not a strict enough criterion to distinguish between Blazhko and non-Blazhko stars. Namely, often many signals (resolved or unresolved) were present in the vicinity of first-overtone and its harmonics. Such signals can come from irregular period changes that mimic quasi-periodic modulation over the span of the observations or instrumental artefacts. This is particularly the case for short time-base K2 data. Consequently, not all low-amplitude sidepeaks and quasi-periodic modulations can be labelled as the Blazhko effect \citep[see a discussion in][and references therein]{molnar2022}. We decided to also check the temporal variability of amplitude and phase. This served as an additional criterion to distinguish between firm Blazhko stars and dubious cases in which, due to the complexity of the Blazhko effect no firm conclusion is possible on a time baseline as short as that of K2 Campaigns. Examples of amplitude and phase changes for two stars are presented in Fig.~\ref{fig:mod}. In the top panel, we plotted a star that was classified as a firm Blazhko star. In the bottom panel, we plotted a star that was marked as a Blazhko star based only on the presence of sidepeaks, but the amplitude and phase variability is too irregular to confirm that.

We also divided Blazhko stars into BL1 and BL2 subclasses. In BL2 stars, we detected sidepeaks at both sides of the dominant frequency. In BL1 stars, sidepeaks on one side of the main peak, so-called dublets are detected.

Besides the additional signals connected to the Blazhko effect or the RR$_{0.61}$ and RR$_{0.68}$ groups, in some stars we detected additional signals that do not fall into the mentioned categories. We present these stars in Sec~\ref{Sec.additional}.

A number of stars from the sample were not straightforward to classify. These stars are discussed individually in Appendix~\ref{Sec.dubious}.

\section{Results}\label{Sec.results}

\subsection{RRc stars}

We classified 452 targets as RRc stars. In Table~\ref{tab:rrc} we provide their coordinates, mean brightness collected in the Ecliptic Plane Input Catalog \citep[EPIC,][]{epic} and the dominant period derived during the analysis. In the case of seven stars in NGC 5897, the information about mean brightness was not available in the input catalogue, but was calculated from the light curve during the analysis (term $A_0$ in Eq.~\ref{eq.series}).

The positions of RRc stars in sky coordinates are plotted in Fig.~\ref{fig:all_coords} with red points. There are 7 stars in our sample which are likely members of the globular cluster NGC~5897. This and the other globular clusters observed in the K2 Campaigns will be studied elsewhere in detail (Kalup et al.\ in prep). 

\begin{figure}
    \centering
    \includegraphics[width=\columnwidth]{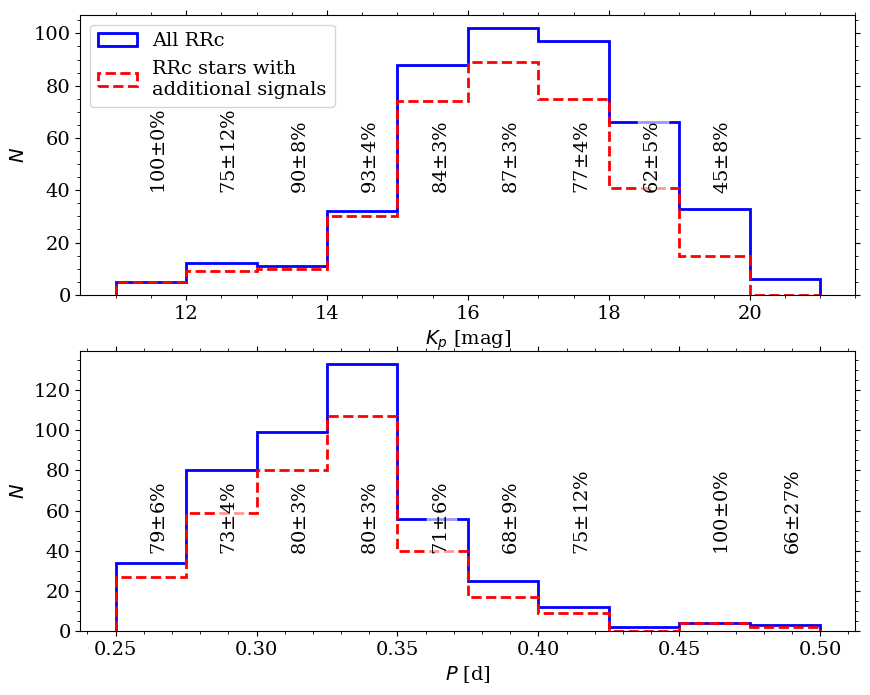}
    \caption{Distribution of observed brightness in $K_p$ band (top panel) and first-overtone period (bottom panel) for stars classified as RRc (blue solid line). With red dashed line we plotted distributions for RRc stars in which we detected additional low-amplitude signals. We also provided incidence rate of RRc stars with the additional signals for each bin of both distributions. The error was calculated assuming Poisson distribution \citep[see e.g.][]{alcock2003}.}
    \label{fig:rrc_hist} 
\end{figure}

In the top panel of Fig.~\ref{fig:rrc_hist} we present the distribution of the observed $K_p$ band brightnesses for RRc stars (blue solid line). The range of brightness in the sample is wide, from 11 mag to 21 mag, with a majority of stars being fainter than 15 mag. The average brightness in the sample is around 16.7 mag.

The distribution of dominant periods for all RRc stars is presented in the bottom panel of Fig.~\ref{fig:rrc_hist} (blue solid line). The shortest and the longest periods in the sample are 0.24\,d and 0.56\,d, respectively. The average period in the sample is 0.33\,d and the majority of stars have periods shorter than around 0.35\,d.

Distributions in Fig.~\ref{fig:rrc_hist} plotted with red dashed lines correspond to RRc stars in which we detected any kind of additional signal. These can be either RR$_{0.61}$, RR$_{0.68}$, Blazhko stars or stars where the additional signals are of unkown origin. All these star will be discussed in detail further in the text. In Fig.~\ref{fig:rrc_hist} we also provided for each bin the incidence rate of RRc stars with additional signals. The incidece rate slightly decreases for decreasing brightness. In particular, we did not detect any additional signals in stars with mean brightness above 20\,mag. We did not notice any significant dependence of the incidence rate on the dominant pulsation period.

\begin{table}[]
    \centering
    \begin{tabular}{ccccc}
    EPIC & RA [deg] & Dec [deg] & $K_p$ & $P$ [d] \\
    \hline
  200194931*   &      229.3261   &      -21.0019   &      15.4057   &      0.498131\\
  200194935*   &      229.3401   &      -21.0344   &      13.7672   &      0.454615\\
  200194936*   &      229.3422   &      -21.019    &      15.937    &      0.419887\\
  200194938*   &      229.3423   &      -20.9855   &      16.8162   &      0.348666\\
  200194944*   &      229.3611   &      -20.9868   &      16.02     &      0.348666\\
  200194945*   &      229.3621   &      -20.9689   &      15.3926   &      0.341787\\
  200194946*   &      229.3771   &      -21.0544   &      15.3212   &      0.332665\\
  201133852    &      183.4337   &      -5.6942    &      16.269    &      0.321896\\
  201150676    &      180.2979   &      -5.2797    &      17.784    &      0.283579\\
  201161411    &      180.336    &      -5.0233    &      16.004    &      0.283133\\
\vdots	&	\vdots	&	\vdots	&	\vdots	&	\vdots	\\
    \end{tabular}
    \caption{Basic properties of stars classified as RRc. Consecutive columns provide an EPIC number of the star, its position, brightness and dominant period of pulsations. A full table is available online.}
    \label{tab:rrc}
\end{table}

\subsection{RR$_{0.61}$ stars}\label{Sec.061}

We found the additional signals corresponding to the RR$_{0.61}$ group in 281 stars. This corresponds to an incidence rate of 62 per cent. Properties of these stars are collected in Table~\ref{tab:061}, where we provided the period of the first overtone, $P_{\rm 1O}$, the period of the additional signal, $P_{0.61}$, their period ratio, the Fourier amplitude of the first overtone, $A_{\rm 1O}$, and the amplitude ratio of the additional signal to the first-overtone amplitude, $A_{0.61}/A_{\rm 1O}$. Remarks are provided in the last column. For some stars, there is more than one row. In these stars, we detected more than one additional signal that forms a period ratio within a range of 0.57--0.65. Detection of several signals is common for RR$_{0.61}$ stars and will be discussed further in this section.

\begin{table*}
    \centering
    \begin{tabular}{lllllll}
EPIC	&	P$_{\rm 1O}$ [d]	&	P$_{\rm 0.61}$ [d] &	P$_{\rm 0.61}$/P$_{\rm 1O}$	&	A$_{\rm 1O}$	& A$_{\rm 0.61}$/A$_{\rm 1O}$	&	Remarks	\\
\hline
200194936 & 0.419874(1)   &  0.25454(1)   &   0.60623  &  0.1135(1)  &  0.023  &    c,s    \\   
          & 0.419874(1)   &  0.25994(2)   &   0.61909  &  0.1135(1)  &  0.011  &    \\
201133852 & 0.388083(2)   &  0.23803(3)   &   0.61335  &  0.2000(2)  &  0.010  &    c,s,0.68  \\
          & 0.388083(2)   &  0.23184(2)   &   0.59740  &  0.2000(2)  &  0.009  &    \\
          & 0.388083(2)   &  0.24456(1)   &   0.63016  &  0.2000(2)  &  0.011  &    \\
201161411 & 0.28072278(8) &  0.172867(3)  &   0.61579  &  0.2253(1)  &  0.005  &    c,s \\      
          & 0.28072278(8) &  0.167785(2)  &   0.59769  &  0.2253(1)  &  0.005  &    \\
          & 0.28072278(8) &  0.170254(3)  &   0.60649  &  0.2253(1)  &  0.003  &    \\
201200243 & 0.3227487(1)  &  0.199739(2)  &   0.61887  &  0.2285(2)  &  0.012  &    c,s,d \\    
          & 0.3227487(1)  &  0.197997(9)  &   0.61347  &  0.2285(2)  &  0.008  &    \\
\vdots	&	\vdots	& \vdots	&	\vdots	&	\vdots	&	\vdots	&	\vdots	\\
    \end{tabular}
    \caption{Properties of RR$_{0.61}$ stars. Consecutive columns provide the star ID, period of the first overtone, period of the additional signal, period ratio, amplitude of the first overtone and amplitude ratio of the additional signal and the first overtone. The last columns provide remarks. The full table is available online. \newline
    Remarks: `c' -- combination signal of the first overtone and the additional signal; `d' -- another additional signal which does not correspond to the RR$_{0.61}$, RR$_{0.68}$ or Blazhko stars; `s' -- subharmonics of the RR$_{0.61}$ signal; `bl' -- the Blazhko effect; `0.68' -- signal corresponding to RR$_{0.68}$ stars.}
    \label{tab:061}
\end{table*}

\begin{figure}
    \centering
    \includegraphics[width=\columnwidth]{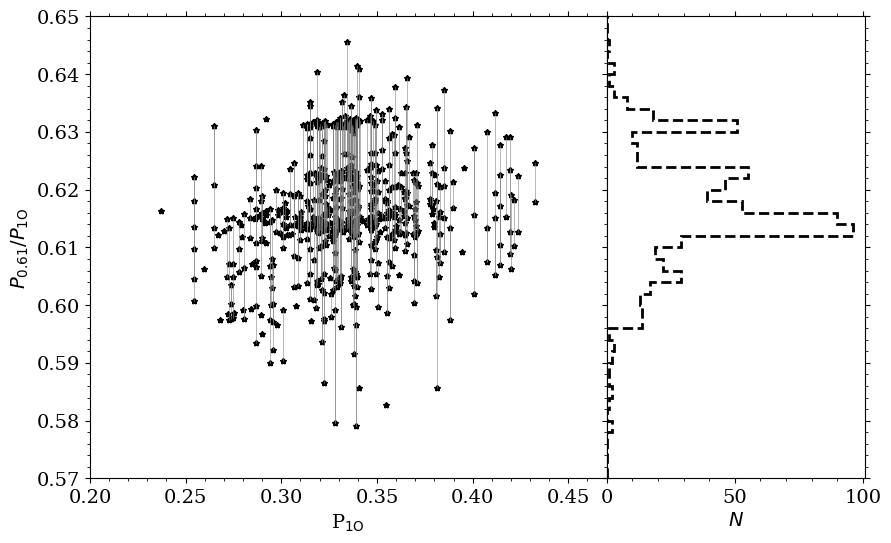}
    \caption{Petersen diagram for RR$_0.61$ stars. Signals detected in the same star are connected with a line. Right panel: histogram of period ratios.}
    \label{fig:pet061}
\end{figure}

In Fig.~\ref{fig:pet061} we plotted a Petersen diagram for RR$_{\rm 0.61}$ stars using all of the additional signals detected in these stars (i.e., all rows from Table~\ref{tab:061}). Signals that correspond to the same star are connected with a line. In the right panel of Fig.~\ref{fig:pet061} we plotted the distribution of period ratios. In the Petersen diagram, our sample forms three already well-known sequences for period ratios around 0.61, 0.62 and 0.63 \citep[see e.g.][]{netzel_census}. Besides these three sequences, there is noticeable scatter towards lower period ratios, below 0.61, where two additional maxima are visible. The first one corresponds to a period ratio of around 0.605 and the second to around 0.598. There are six stars with even lower values, i.e. below 0.59. In four of them, the signal that forms the low period ratio is present together with signals forming period ratios in the expected range of 0.60 -- 0.64. 

Two stars in which the only detected signal forms low period ratio are EPIC\,220654797 with $P_{0.61}/P_{\rm 1O} \approx 0.5857$ and EPIC\,250003296 with $P_{0.61}/P_{\rm 1O} \approx 0.58273$. Such values would also fit double-mode pulsations in the fundamental mode and second overtone. We checked the Fourier decomposition parameters for both stars. The position of EPIC\,220654797 in the Fourier decomposition parameters plot (Fig.~\ref{fig:fdp}) is in between the groups of RRab and RRc stars. The classification of this star as pulsating in the fundamental mode and second overtone is therefore also possible. In the case of EPIC\,250003296, its position in Fig.~\ref{fig:fdp} is very typical for RRc stars. In this star, however, we did not detect any combination signals between the additional periodicity and first overtone, so there is a strong probability that the observed signal originates from blending. In both outlying stars, we detected the Blazhko effect.

\begin{figure}
    \centering
    \includegraphics[width=\columnwidth]{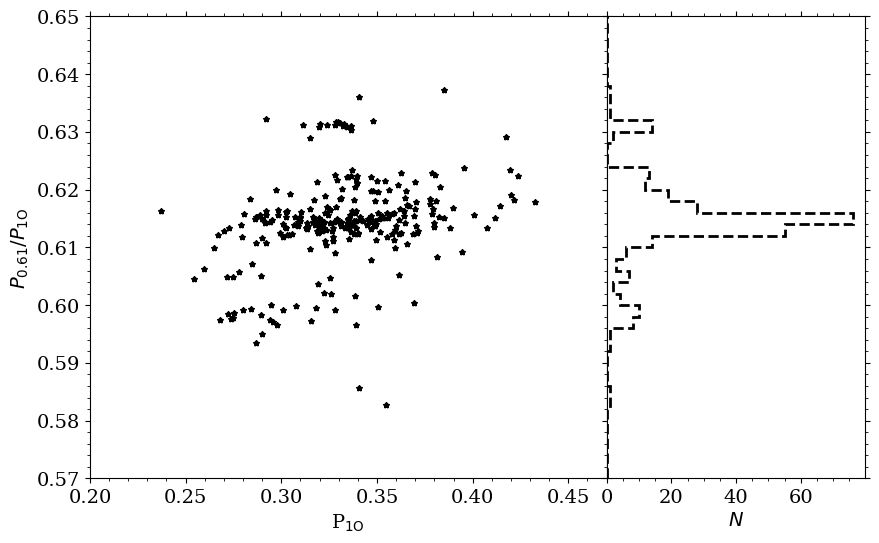}
    \caption{Same as Fig.~\ref{fig:pet061}, but only for the single additional signal of the highest amplitude.}
    \label{fig:pet061single}
\end{figure}

As already visible from the Petersen diagram in Fig.~\ref{fig:pet061}, we detected more than one additional signal in the majority of stars. In fact,  we detected  a single signal in only 95 stars that has $P_{0.61} \in \left\langle 0.57, 0.65 \right\rangle P_{\rm 1O}$. In Fig.~\ref{fig:pet061single} we plotted the Petersen diagram for RR$_{0.61}$ stars using the only detected signal or, in the case of stars with multiple signals detected, the additional signal with the highest amplitude. A histogram of period ratios is presented in the right panel of Fig.~\ref{fig:pet061single}. The most populated sequence is still located at a period ratio of around 0.61. There is also significant scatter around this sequence. Within the range of this scatter we can see a hint of a sequence centred at a period ratio of 0.62. The sequence located at period ratios of 0.63 is also well visible. There is significantly less scatter around this sequence. Interestingly, we clearly see the new sequence formed in the Petersen diagram at a period ratio of around 0.60.
Three sequences located at period ratios of 0.598, 0.61 and 0.63 are also clearly visible in the distribution of period ratios in the right panel of Fig.~\ref{fig:pet061single}. The sequences located at period ratios of around 0.605 and 0.62 are still visible, but less pronounced.

\begin{figure}
    \centering
    \includegraphics[width=\columnwidth]{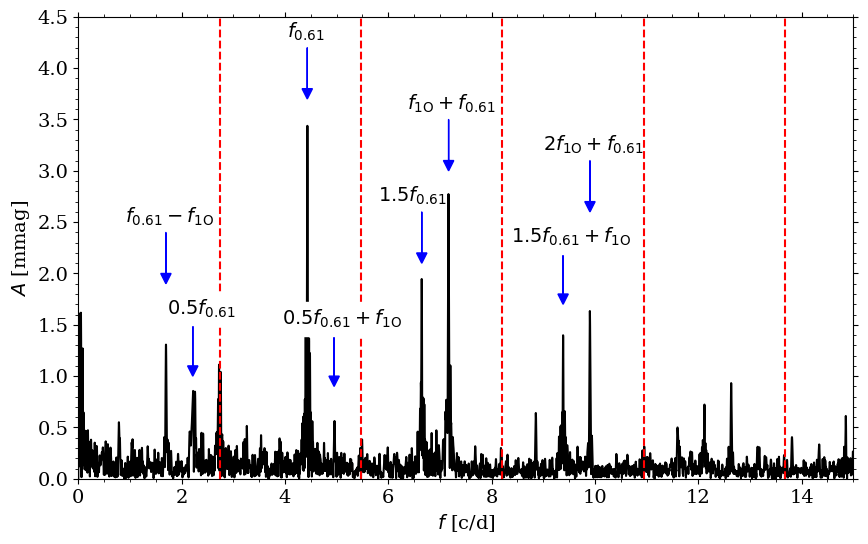}
    \caption{Frequency spectrum for EPIC\,228763070 after prewhitening with the first overtone and its harmonics. Positions of prewhitened signals are marked with red dashed lines. The highest additional signals present in the frequency spectrum are marked with arrows and labelled.}
    \label{fig:061example_fs}
\end{figure}

\begin{figure}
    \centering
    \includegraphics[width=\columnwidth]{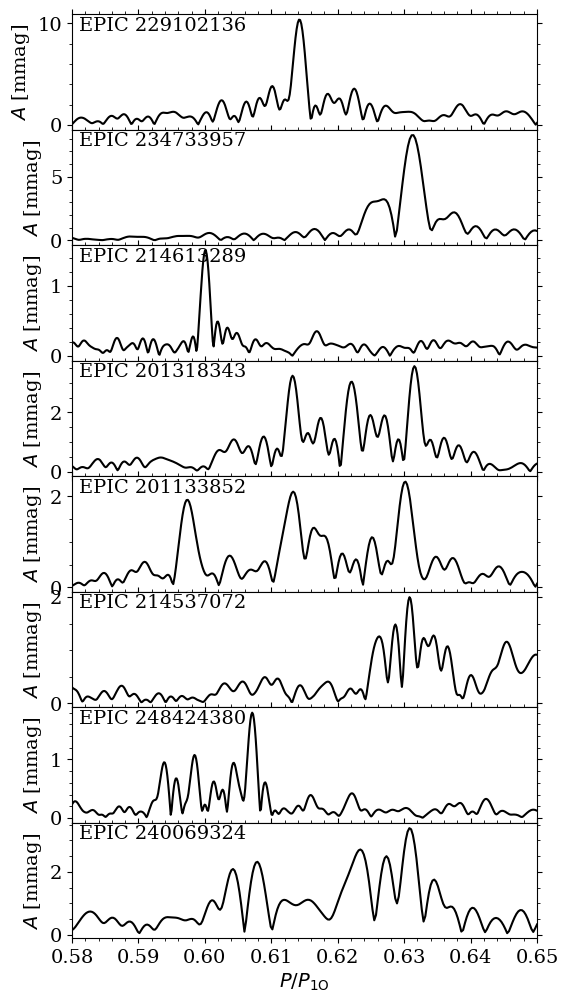}
    \caption{Examples of frequency spectra for eight RR$_{0.61}$ stars for a frequency range corresponding to $f_{0.61}$ signals. The horizontal axis is scaled with the first-overtone periods. Star's ID is provided in the upper left corner of each panel.}
    \label{fig:061examples}
\end{figure}

\begin{figure*}
    \centering
    \includegraphics[width=\textwidth]{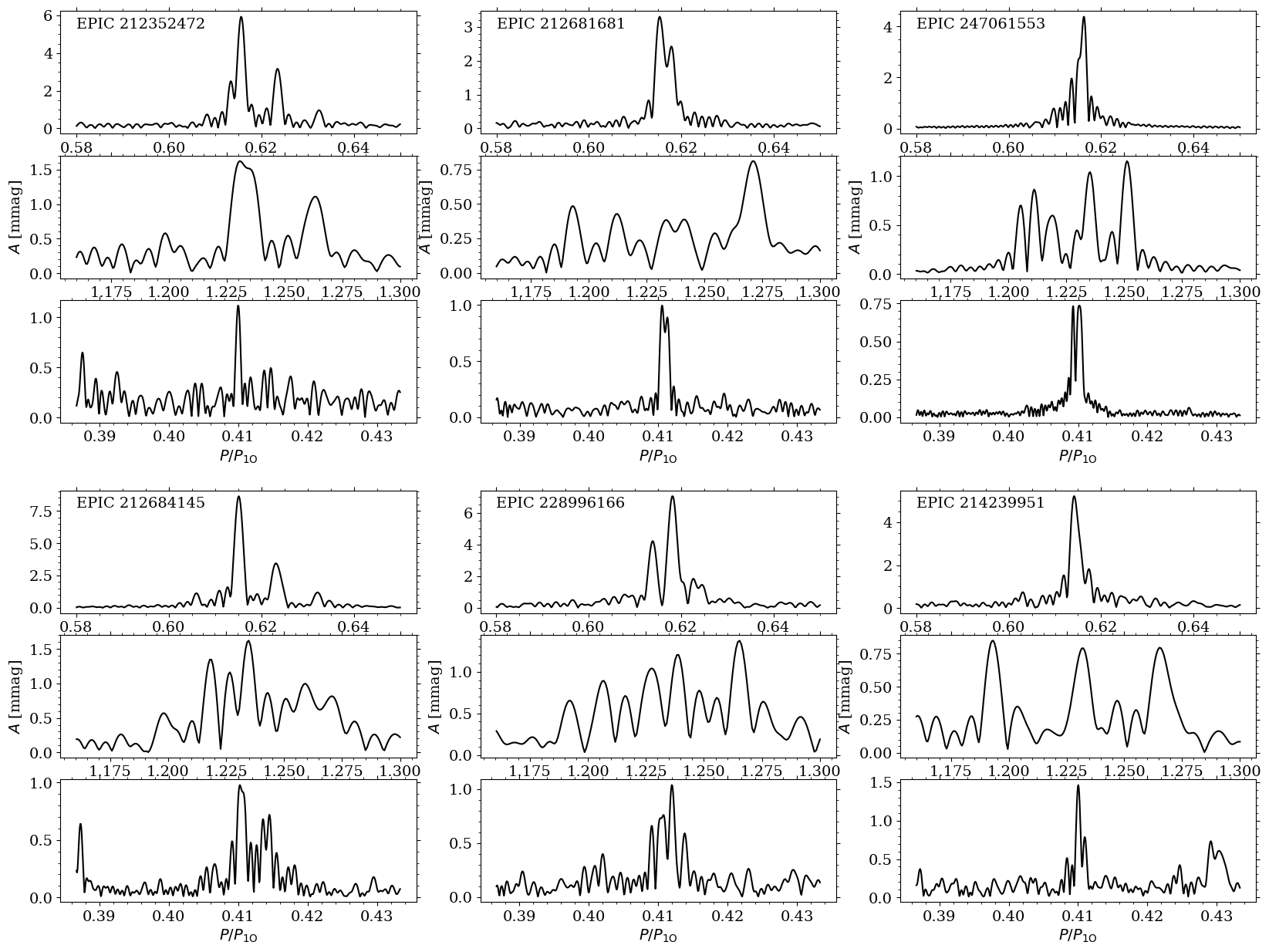}
    \caption{Frequency spectrum examples for six RR$_{0.61}$ stars. For each star, the top, middle and lower panels show the frequency ranges corresponding to the $f_{0.61}$, 0.5$f_{0.61}$, and 1.5$f_{0.61}$ signals, respectively. The horizontal axis is scaled with the first-overtone periods. EPIC numbers are provided in the upper right corner of the top panel.}
    \label{fig:061example_subharm}
\end{figure*}

In Fig.~\ref{fig:061example_fs} we show an example of a frequency spectrum for an RR$_{0.61}$ star, EPIC\,228763070, after prewhitening with the first overtone and its harmonics, whose positions are marked with red dashed lines. A large number of additional signals remain after prewhitening. The highest signal corresponds to the $f_{0.61}$ signal. Both signals corresponding to the subharmonics, i.e. signals at $0.5f_{0.61}$ and $1.5f_{0.61}$ are detected. Such signals are common for RR$_{0.61}$ stars, especially in space-based photometry. Stars in which we detected subharmonics are marked with `s' in the remarks column of Table~\ref{tab:061}. We also detected signals arising due to linear combinations between $f_{0.61}$, its subharmonics and first-overtone frequency. Stars in which we detected combination signals between $f_{0.61}$ and first overtone are marked with `c' in the remarks column of Table~\ref{tab:061}.

In Fig.~\ref{fig:061examples} we plotted frequency spectra of eight of RR$_{0.61}$ stars as examples for the variety seen in additional signals. Stars' EPIC numbers are provided in the top left corner of each panel. We show the frequency range corresponding to the $f_{0.61}$ signals only and scale the horizontal axis with the first-overtone period, so the positions of signals can be directly compared with the positions of the stars in the Petersen diagram (Fig.~\ref{fig:pet061} and Fig.~\ref{fig:pet061single}). In the top three panels, we show examples of stars where only one additional signal was detected. Here we chose stars that belong to three different sequences in the Petersen diagram. Specifically, they are members of the 0.61, 0.63 and 0.60 sequences, respectively, starting from the top panel. As already shown, more than one signal was detected in this frequency range in the majority of RR$_{0.61}$ stars. The next panels of Fig.~\ref{fig:061examples} show examples of such stars. In the fourth and fifth panels, three signals were detected. These signals are well resolved from each other and form period ratios that fit the three sequences in the Petersen diagram well simultaneously. In the case of 201318343 (fourth panel), the star belongs to the 0.61, 0.62 and 0.63 sequences. In the case of EPIC\,201318343 (fifth panel), the star belongs to 0.60, 0.61 and 0.63 sequences. In the last three panels, we present examples of stars in which many additional signals were detected, not only those that correspond directly to well-defined sequences in the Petersen diagram.

The subharmonics of the additional signals, i.e. signals at the half-integer frequencies, are often observed in RR$_{0.61}$ stars \citep[see e.g.][]{netzel_census,moskalik2015}. From the ground-based photometry, subharmonics at 0.5$f_{0.61}$ are detected most of the time. In the space-based photometry, both subharmonics at 0.5$f_{0.61}$ and 1.5$f_{0.61}$, respectively, are reported. We found signals at subharmonic frequencies in 183 stars, which corresponds to 65 per cent of the RR$_{0.61}$ stars. Stars with detected subharmonics are marked in Table~\ref{tab:061} with `s' in the remarks columns. In 104 stars we detected both subharmonics at 0.5$f_{0.61}$ and 1.5$f_{0.61}$. In 40 stars we detected subharmonics only at 0.5$f_{0.61}$ and in 39 stars we detected subharmonics only at 1.5$f_{0.61}$.

\begin{figure}
    \centering
    \includegraphics[width=\columnwidth]{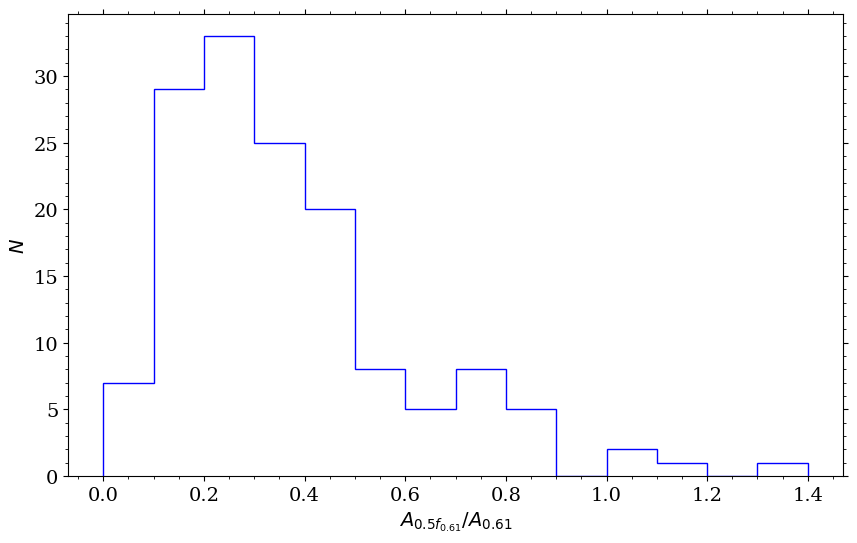}
    \caption{Distribution of amplitude ratio of subharmonic at $0.5f_{0.61}$ and signal at $f_{0.61}$.}
    \label{fig:subh_amp}
\end{figure}

Typically, the signals corresponding to the subharmonics show temporal variations in amplitudes and phases \citep[see][]{moskalik2015}. In the frequency spectra, this variability manifests as wide signals, often located offset from the exact frequencies at 0.5$f_{0.61}$ or 1.5$f_{0.61}$. In Fig.~\ref{fig:061example_subharm} we show six examples of stars with both subharmonics to present the variety of these signals in the frequency domain. The structure formed by the subharmonics at 0.5$f_{0.61}$ is much more complex than the $f_{0.61}$ signal itself or the subharmonics at 1.5$f_{0.61}$. Subharmonics at 0.5$f_{0.61}$ typically form an extended section of power excess over a wide frequency range. Interestingly, the subharmonics at 1.5$f_{0.61}$ are significantly narrower than their low-frequency counterparts, and show more resemblance to signals at $f_{0.61}$. According to the explanation by \cite{dziembowski2016}, signals at 0.5$f_{0.61}$ are the true frequencies of non-radial modes excited in these stars, while signals observed at $f_{0.61}$ and 1.5$f_{0.61}$ are harmonics of those non-radial modes. Due to the cancellation effect, signals at $f_{0.61}$ have typically higher amplitudes than subharmonics at 0.5$f_{0.61}$, i.e., the parent frequencies. However, this is not always the case. In Fig.~\ref{fig:subh_amp} we plotted the distribution of amplitude ratios of the 0.5$f_{0.61}$ and $f_{0.61}$ signals. The subharmonics at 0.5$f_{0.61}$ typically have much smaller amplitudes than the signals at $f_{0.61}$: the amplitude ratio is smaller than 0.5 for the majority of stars. But in four stars the amplitude of the subharmonics is higher. The ratio of these stars is lower than in the Galactic bulge sample, where this was the case for 35 out of 114 RR$_{0.61}$ stars with subharmonics \citep{netzel_census}.

\subsection{RR$_{0.68}$ stars}

We found the additional signal corresponding to the RR$_{0.68}$ group in 67 stars, giving us an incidence rate of 15 per cent. The properties of these stars are collected in Table~\ref{tab:068}. We provided the star IDs, period of the first overtone, $P_{\rm 1O}$, and the additional signal, $P_{0.68}$, their ratio, the amplitude of the first overtone, $A_{\rm 1O}$, and the ratio of the amplitude of the additional signal and first-overtone amplitude, $A_{0.68}/A_{\rm 1O}$. In the last column, we provided remarks. As opposed to the RR$_{0.61}$ stars, the additional signals in RR$_{0.68}$ stars are typically coherent and single peaks. However, in three stars we detected more than one signal that forms a period ratio close to 0.68. In Table~\ref{tab:068} there is more than one row for these stars.

\begin{figure}
    \centering
    \includegraphics[width=\columnwidth]{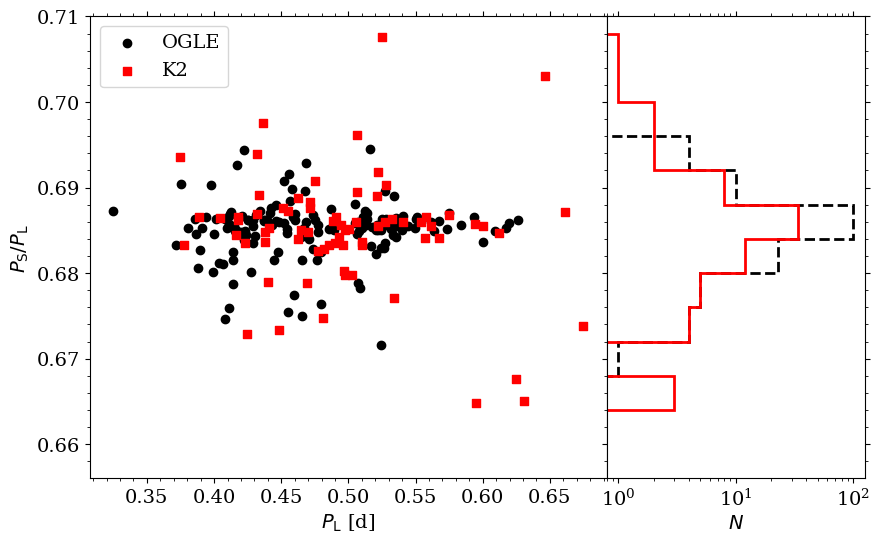}
    \caption{Petersen diagram for RR$_{0.68}$ stars. Stars detected in the K2 data are plotted with red squares. Stars from the OGLE sample are plotted with black circles. Right panel: histogram of period ratios for the two samples.}
    \label{fig:pet068}
\end{figure}

The RR$_{0.68}$ stars are plotted in the Petersen diagram in Fig.~\ref{fig:pet068}. For reference, we also plotted RR$_{0.68}$ stars detected in the OGLE data \citep{netzel068}. The stars studied here are centred on a period ratio of around 0.685, which is slightly lower than what was observed for the OGLE sample, where the average period ratio was around 0.686. The majority of stars have period ratios ranging from 0.678 to 0.692, with five stars having higher, and five other stars having lower ratios, respectively. The highest period ratio observed in the sample is 0.70779, for EPIC\,228757559. In this star, we also detected the $f_{0.61}$ signal. Another outlier with a high period ratio is EPIC\,200194935. In this star we see three additional signals in that have $f_{0.68} \in \left\langle 0.66, 0.71\right\rangle$. The signal of the highest amplitude forms a period ratio that fits well the RR$_{0.68}$ group, $P_{\rm 1O}/P_x \approx 0.687$. The second-highest amplitude signal forms a slightly lower period ratio, around 0.673. The signal with the highest period ratio has the lowest amplitude of the three, its signal-to-noise ratio being only 4.16. The lowest period ratio in the sample is 0.66764, observed in EPIC\,246217239. The detection in this star is significant with a signal-to-noise ratio of around 7.9. In this star, we also detected a $f_{0.61}$ signal and an additional signal with low-amplitude that fits the RRd sequence in the Petersen diagram. Interestingly, the majority of stars that are outliers from the RR$_{0.68}$ group in the Petersen diagram have additional signals with periods $P_{0.68}>0.5$\,d.

\begin{table*}
    \centering
    \begin{tabular}{lllllll}
EPIC	&	P$_{\rm 1O}$ [d]	&	P$_{\rm 0.68}$ [d] &	P$_{\rm 1O}$/P$_{\rm 0.68}$	&	A$_{\rm 1O}$ [mag]	& A$_{\rm 0.68}$/A$_{\rm 1O}$	&	Remarks	\\
\hline
200194935 & 0.45461508(1) &  0.661553490(10) & 0.68719  &  0.01075(2) &  0.036    &  c   \\     
          & 0.45461508(1) &  0.674741940(10) & 0.67376  &  0.01075(2) &  0.026    &  \\
          & 0.45461508(1) &  0.64662953(9)   & 0.70305  &  0.01075(2) &  0.011    &  \\
200194938 & 0.34866609(1) &  0.51002570(2) &  0.68362   & 0.02574(10) & 0.039     &    \\       
200194944 & 0.34866608(1) &  0.51026228(3) &  0.68331   & 0.0358(1)   & 0.044     &      \\     
200194945 & 0.34178674(1) &  0.502827450(10) & 0.67973  &  0.02629(3) &  0.028    &      \\     
201133852 & 0.388083(2)   &  0.5673(2)     &  0.68408   & 0.2000(2)   & 0.005     & 0.61   \\   
201552850 & 0.341037(2)   &  0.49782(5)    &  0.68507   & 0.0776(2)   & 0.060     & c,bl,d \\
\vdots & \vdots	&	\vdots	&	\vdots	&	\vdots	&	\vdots	\\
    \end{tabular}
    \caption{Properties of $RR_{0.68}$ stars. Consecutive columns provide the star EPIC number, period of the first overtone, period of the additional signal, period ratio, amplitude of the first overtone and amplitude ratio of the additional signal and the first overtone. The last columns provide remarks. The full table is available online. \newline
    Remarks: `c' -- combination signal of the first overtone and the additional signal (two stars marked with `c?' are discussed in the Appendix); `d' -- another additional signal which does not correspond to the RR$_{0.61}$, RR$_{0.68}$ or Blazhko stars; `bl' -- the Blazhko effect; `0.61' -- signals corresponding RR$_{0.61}$ stars; `h' -- harmonic of the additional signal detected}
    \label{tab:068}
\end{table*}

\begin{figure}
    \centering
    \includegraphics[width=\columnwidth]{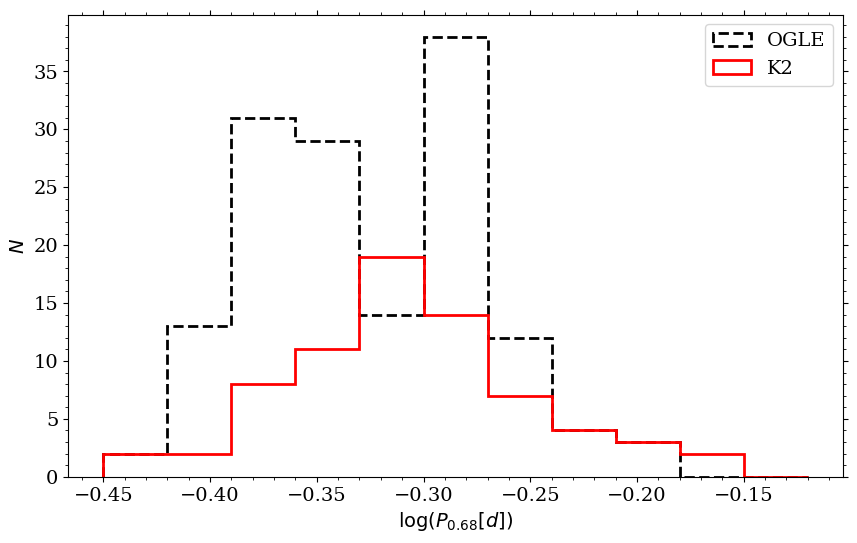}
    \caption{Histogram of periods of the additional signal for RR$_{0.68}$ stars detected in the K2 data (red solid line) and in the OGLE data (black dashed line).}
    \label{fig:p_hist}
\end{figure}

In Fig.~\ref{fig:p_hist} we plotted a histogram of the $P_{0.68}$ periods. For comparison, we included the sample of RR$_{0.68}$ stars based on the OGLE data. The highest number of stars in the K2 sample corresponds to periods of around 0.5\,d. Interestingly, in the distribution for the OGLE sample, there is a minimum for $P_{0.68} \approx 0.5$\,d. In ground-based data, as in the case of the OGLE data, periods around 0.5\,d correspond to the position of possible instrumental signals or daily aliases of long trends. Consequently, the distribution of the K2 sample is likely more representational for the RR$_{0.68}$ group.

In 32 RR$_{0.68}$ stars we found signals corresponding to the RR$_{0.61}$ group. These stars are marked with `0.61' in the remarks column of Table~\ref{tab:068} and with `0.68' in the remarks column of Table~\ref{tab:061}. These stars are particularly interesting in the context of simultaneous theoretical explanations of both types of modes (see discussion in Sec.~\ref{sec:disc_0.68}) So far, only six such stars were known. One star was observed in the original Kepler field \citep{moskalik2015}, three stars were detected in the OGLE sample \citep{netzel_census}, and two in the TESS observations \citep{molnar2022}. Here we significantly increased the number of known stars that show both kinds of additional signals.

\begin{figure}
    \centering
    \includegraphics[width=\columnwidth]{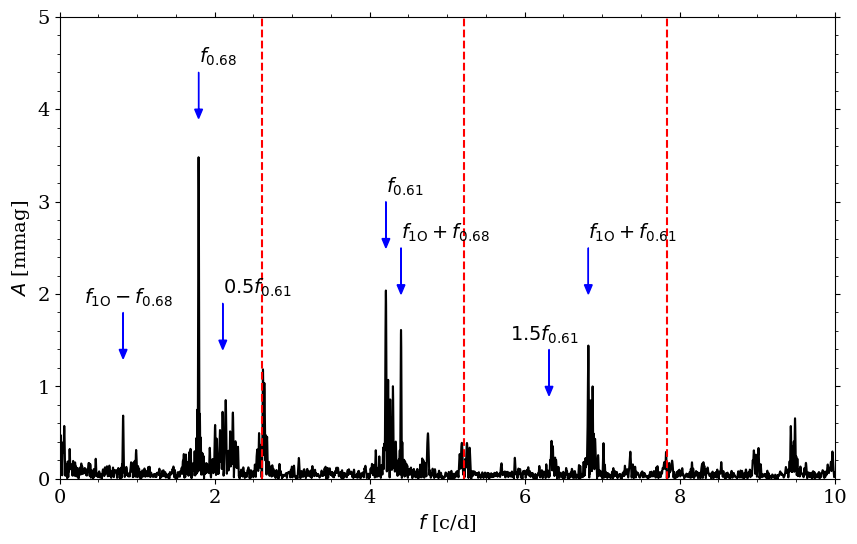}
    \caption{Frequency spectrum for EPIC\,249784033 after prewhitening with the first overtone and its harmonics (marked with red dashed lines). The most important signals are marked with arrows and labelled.}
    \label{fig:061_068}
\end{figure}

In Fig.~\ref{fig:061_068} we plotted the frequency spectrum of EPIC\,249784033 which is one of the stars that show the $f_{0.68}$ and the $f_{0.61}$ signals simultaneously. The positions of the prewhitened first overtone and its harmonics are marked with red dashed lines. The highest and most important signals visible in the frequency spectrum are marked with arrows and described in the figure. The highest peak corresponds to the $f_{0.68}$ signal. It also forms a combination with the first overtone, $f_{\rm 1O}+f_{0.68}$, which is located close to the $f_{0.61}$ signal. However, both are separated and well resolved. We did not detect a combination signal between $f_{0.61}$ and $f_{0.68}$.

In 14 of RR$_{0.68}$ stars we found Blazhko modulation. These stars are marked with `bl' in the remarks column of Table~\ref{tab:068}. Blazhko stars will be discussed in detail in Sec.~\ref{Sec.blazhko}.

Beside the dominant first-overtone signal, the RR$_{0.68}$ signal, its combination and harmonics, we detected additional signals of mostly unknown origin in 28 RR$_{0.68}$ stars. These stars are marked with `d' in the remarks column of Table~\ref{tab:068}.

\subsection{Blazhko stars}\label{Sec.blazhko}

\begin{figure*}
    \centering
    \includegraphics[width=\textwidth]{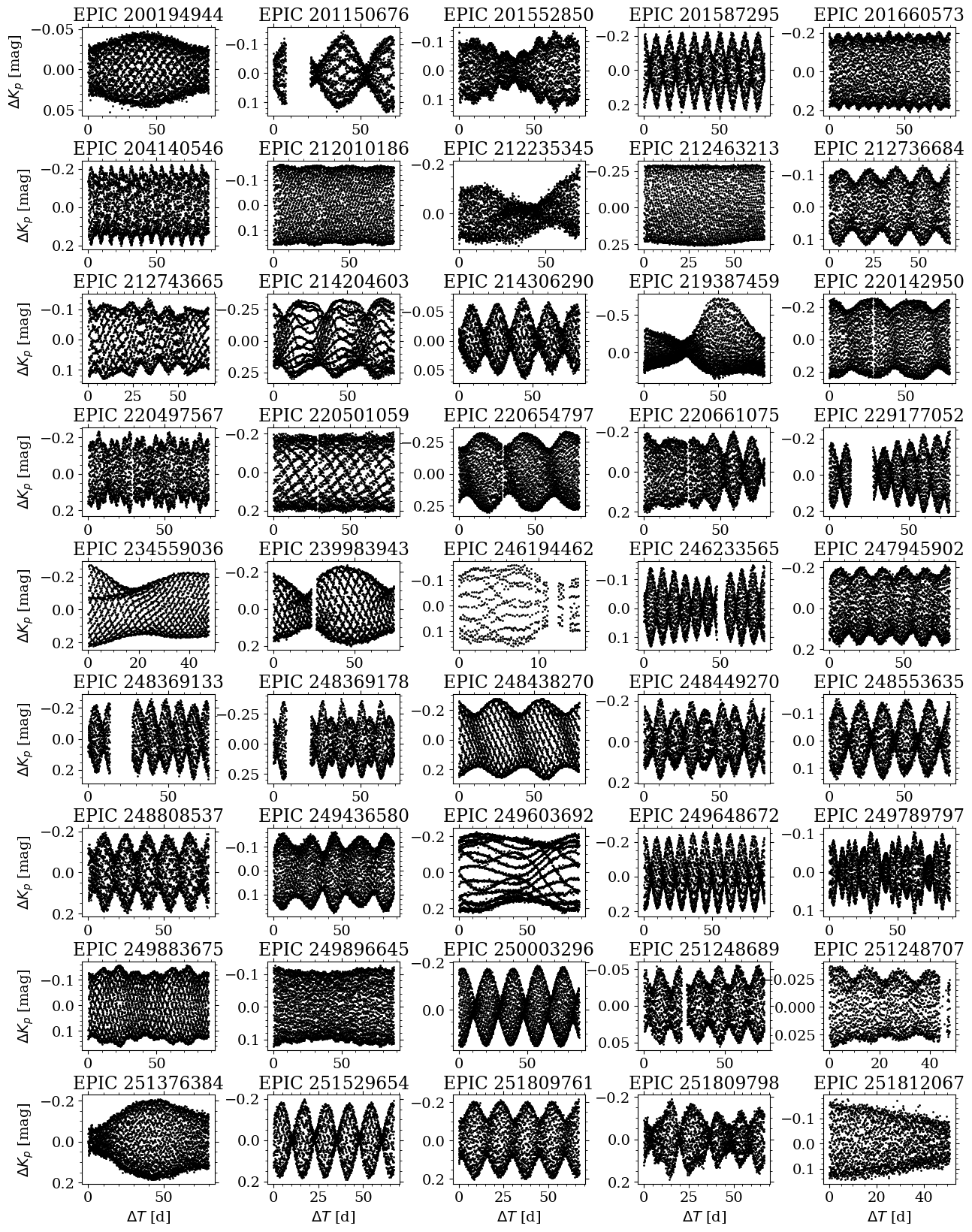}
    \caption{RRc stars with visible Blazhko modulation in the light curve. }
    \label{fig:bl_gallery}
\end{figure*}

We detected Blazhko modulation based on visual inspection of light curves in 45 stars. Their light curves are presented in Fig.~\ref{fig:bl_gallery}, with EPIC numbers included above each panel. The diversity of modulation periods and amplitudes is visible in Fig.~\ref{fig:bl_gallery}. Particularly interesting are stars that show multi-periodic modulation, e.g. EPIC\,229177052, EPIC\,201552850. Another interesting object is EPIC\,220661075 (third row and third column in Fig.~\ref{fig:bl_gallery}) in which the period and amplitude of modulation seem to be changing during the 80\,d observing period. In the case of several stars, the data length is too short to cover the whole modulation period. In the frequency spectra of such stars, the modulation would show up as signals unresolved from the dominant frequency and its harmonics. The data length of the K2 data is around 80 days. This means that with the adopted criterion for resolution in the frequency spectrum, the longest Blazhko period possible to detect is around 40\,d. Only those stars are the topic of further analysis of the Blazhko effect.

In Table~\ref{tab:bl} we collected stars in which the Blazhko modulation manifests as resolved multiplets in the frequency spectrum. This is the case for 57 stars, which gives the incidence rate of the Blazhko effect in the RRc K2 sample of 12.6 per cent. We note, that some stars classified as Blazhko are not presented in Fig.~\ref{fig:bl_gallery}, because the amplitude of modulation is too small to be visible in the lightcurve. In eleven more stars, we see clear modulation of the light curve, but the modulation period is too long for the available data. Including these stars, the number of Blazhko stars increases to 68, which corresponds to the incidence rate of 15 per cent.

\begin{table*}
    \centering
    \begin{tabular}{lllllll}
EPIC	&	P$_{\rm 1O}$ [d]	&	P$_{\rm B}$ [d]	&	A$_{\rm 1O}$ [mag]	&	A$_+$ [mag]	&	A$_-$ [mag]	&	Remarks	\\
\hline
201150676 & 0.2650824(2)  &  27.854(3)    &   0.0730(2)  &              & 0.0489(2)  &  d,m,v     \\
201383569 & 0.30790309(2) &  13.512(4)    &   0.22899(8) &  0.00246(8)  & 0.00264(8) &  0.61     \\ 
201544130 & 0.32246555(1) &  14.64788(3)  &   0.18567(7) &  0.00258(7)  & 0.00075(9) &  0.61     \\ 
201552850 & 0.341037(2)   &  10.80(1)     &   0.0776(2)  &  0.0127(2)   & 0.0020(2)  &  0.68,v,d \\ 
201587295 & 0.29281757(1) &  8.3790163(1) &   0.1400(3)  &  0.0174(2)   & 0.0639(3)  &  d,m,v,a  \\ 
201660573 & 0.25984644(1) &  6.7018092(2) &   0.1808(1)  &  0.0008(1)   & 0.0133(1)  &  d,v,0.61 \\ 
203480423 & 0.21414930(9) &  9.011(6)     &   0.16477(5) &  0.00345(5)  &            &           \\ 
204140546 & 0.3103061(3)  &  7.3693(4)    &   0.15466(8) &  0.05479(8)  & 0.04319(8) &  a,d,m,v,0.68 \\
204908361 & 0.31871896(8) &  13.656(5)    &   0.17065(8) &              & 0.00284(8) &  0.61,d \\ 
\vdots	&	\vdots	& \vdots	&	\vdots	&	\vdots	&	\vdots	&	\vdots	\\
    \end{tabular}
    \caption{Properties of Blazhko RRc stars. Consecutive columns provide the EPIC number of the star, period of the first overtone, period of the Blazhko effect, Fourier first-overtone amplitude and amplitudes of sidepeaks on the higher-frequency side ($A_+$) and the lower-frequency side ($A_-$). The last column gives remarks. Note, that for some stars with multiple Blazhko periods there is more than one row in the table. A full table is available online. \newline Meaning of the remarks in the last columns: \newline a - modulation signal in the low-frequency range; c - combination signals between the Blazhko sidepeaks in the case of multiple Blazhko modulations; d - additional signal in the frequency spectrum; m - higher multiplets; 0.61 - additional signal corresponding to RR$_0.61$ stars; 0.68 - additional signal corresponding to RR$_0.68$ stars; v - modulation visible in the light curve data.}
    \label{tab:bl}
\end{table*}

The 57 Blazhko stars for which the modulation period is within the resolution of the Fourier transform are collected in Table~\ref{tab:bl}. Consecutive columns provide the EPIC number of the star, first-overtone and Blazhko periods, first-overtone amplitude and amplitude of sidepeaks at the higher frequency than the first-overtone frequency, $A_+$, and at the lower frequency, $A_-$. The last column provides additional remarks. For some stars, there is more than one row. In these stars, we detected more than one modulation period. We detected two Blazhko periods in four stars, three Blazhko periods in three stars and in one star we found four Blazhko periods. For some stars with multiple modulation periods, we found combinations between the sidepeaks. These stars are marked with `c' in the remarks column of Table~\ref{tab:bl}. In some of the stars it was also possible to detect the signal in the low-frequency range that corresponds to the modulation of mean brightness. Such stars are marked with `a' in the remarks. We marked these stars only when the detection of a signal was clear and not subjected to confusion with long-term instrumental trends. It was typically the case, because we considered only relatively short-period Blazhko stars. However, in the case of some stars, the strong instrumental trend was present in the data and it was not removed completely with spline function. Then, the noise-level in the low-frequency range was raised and might hamper the detection of signals connected to the mean-brightness modulation. In some stars we found multiplets, i.e. sidepeaks in the form of $nf_{\rm 1O} \pm kf_B$, where $k>1$. These stars are marked with `m' in the remarks column.

\begin{figure}
    \centering
    \includegraphics[width=0.5\textwidth]{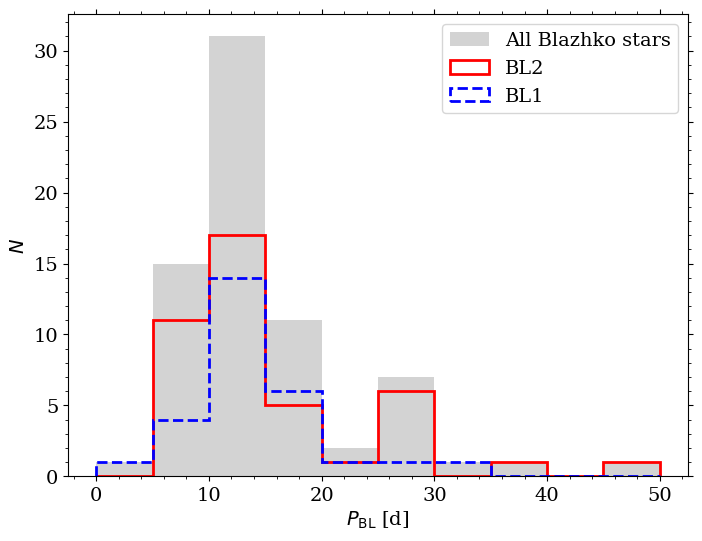}
    \caption{Histogram of the Blazhko periods.}
    \label{fig:hist_pbl}
\end{figure}

In Fig.~\ref{fig:hist_pbl} we plotted a histogram of Blazhko periods from the sample in Table~\ref{tab:bl}. In the case of stars with multiple modulation periods, we used all detected periods. We also compare the distributions of BL1 and BL2 stars. Two maxima are visible in the distribution of all Blazhko stars. The highest maximum corresponds to Blazhko periods of 10--15 days. This maximum is also visible in the period distributions of both the BL1 and BL2 stars, respectively. The second maximum is significantly lower and corresponds to Blazhko periods of 25--30 days. This peak is not visible in the distribution of BL1 stars. The majority of stars have modulation periods between 5 and 20 days. The shortest Blazhko period in the sample is 2.81 days. It was detected in star EPIC\,249789797, among four different Blazhko periods. This Blazhko period also corresponds to the BL1 type, i.e. it does not form a full multiplet in a frequency spectrum. The shortest modulation period detected in a star with a single modulation cycle is 5.57 days and was found in EPIC\,249605297, a BL2 star.

The longest detected Blazhko period is limited by the data length: we found it to be 49\,d, detected in EPIC\,249883675. 
Two more modulation periods were detected in this star. The longest detected period in a star with a single modulation period is 38\,d in EPIC\,212767731, a BL2-type star. As shown in Fig.~\ref{fig:bl_gallery} there are stars in the sample with significantly longer modulation periods, for which we cannot determine the cycle length.

\begin{figure}
    \centering
    \includegraphics[width=0.5\textwidth]{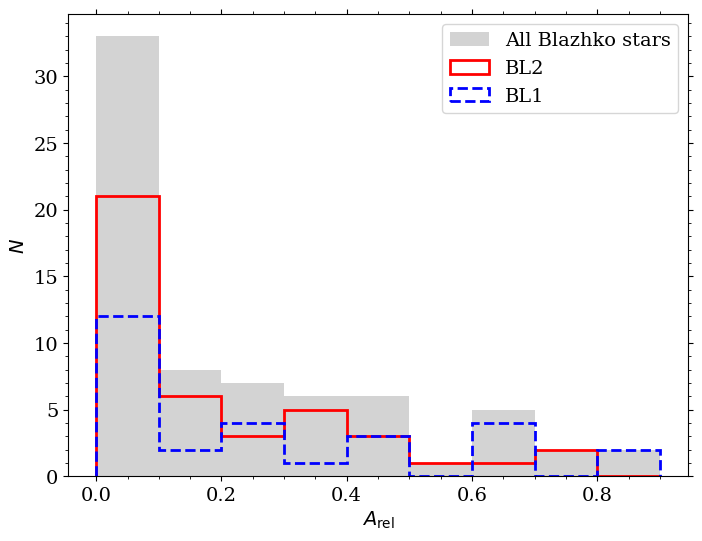}
    \caption{Histogram of the relative modulation for stars with triplets or multiplets detected (BL2) or dublets (BL1).}
    \label{fig:hist_amp_bl}
\end{figure}

In Fig.~\ref{fig:hist_amp_bl} we plotted the relative modulation amplitudes, defined as the ratio of the amplitude of the highest sidepeak to the first-overtone amplitude:
\begin{equation}
    A=\frac{\max(A_+; A_-)}{A_{\rm 1O}}.
\end{equation}
In the case of BL1 stars, we used the amplitude of the only detected sidepeak. In Fig.~\ref{fig:hist_amp_bl} we compared distributions of BL2 and BL1 stars with the distribution of all Blazhko stars. For both groups, the highest number of stars corresponds to the low relative modulation amplitude, below 0.1. Nevertheless, there are several stars in both groups with higher relative modulation amplitude values. The largest value of 0.86 was detected in a BL1 star, EPIC\,251529654. We note, that the frequency content of EPIC\,251529654 is puzzling and it might be in fact an object similar to V37 discussed by \cite{smolec2017} and OGLE-BLG-RRLYR-11754 of OGLE-BLG-RRLYR-15059 discussed by \cite{netzel_census}. Interestingly, there seems to be a gap in the modulation amplitude distribution at around 0.5--0.6. However, the small number of stars with high relative modulation amplitudes makes this observation uncertain.

\begin{figure}
    \centering
    \includegraphics[width=0.5\textwidth]{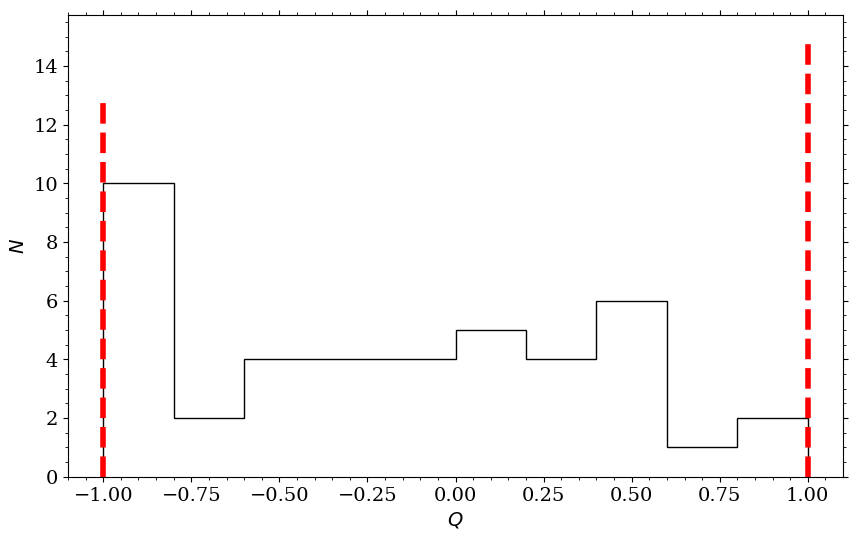}
    \caption{The distribution of the Q parameter (see Eq.~\ref{eq:q_param}) for BL2 stars (black solid line). BL1 stars are plotted with red dashed lines (see text for details).}
    \label{fig:hist_q}
\end{figure}

In Fig.~\ref{fig:hist_q} we plotted the distribution of the asymmetry parameter Q defined as:

\begin{equation}
    Q=\frac{A_+-A_-}{A_++A_-}
    \label{eq:q_param}
\end{equation}

by \cite{alcock2003}. We plotted this distribution in Fig.~\ref{fig:hist_q} with the black line. In the case of BL1 stars, we assigned $Q=+1$ when the sidepeak is located on the higher frequency side or $Q=-1$ in the opposite case. These are plotted with red dashed lines in Fig.~\ref{fig:hist_q}. Interestingly, there are many BL2 stars where the Q parameter is lower than $-0.8$, which means that the sidepeak on the lower frequency side has significantly higher amplitude than the sidepeak on the higher frequency side.

In general, 24 BL2 stars have $Q<0$, whereas 18 BL2 stars have $Q>0$. However, if we consider only stars that have $Q \in \langle -0.8,0.8 \rangle$, then the distribution becomes nearly flat, with 14 stars showing dominant lower-frequency sidepeaks and 16 higher-frequency sidepeaks, respectively. In the case of BL1 stars, there is also no preference for which sidepeak being present. 13 BL1 stars have sidepeaks on the lower-frequency side, whereas 15 stars have sidepeaks on the higher-frequency side.

\subsection{Other additional signals in RRc stars}\label{Sec.additional}

\begin{figure}
    \centering
    \includegraphics[width=\columnwidth]{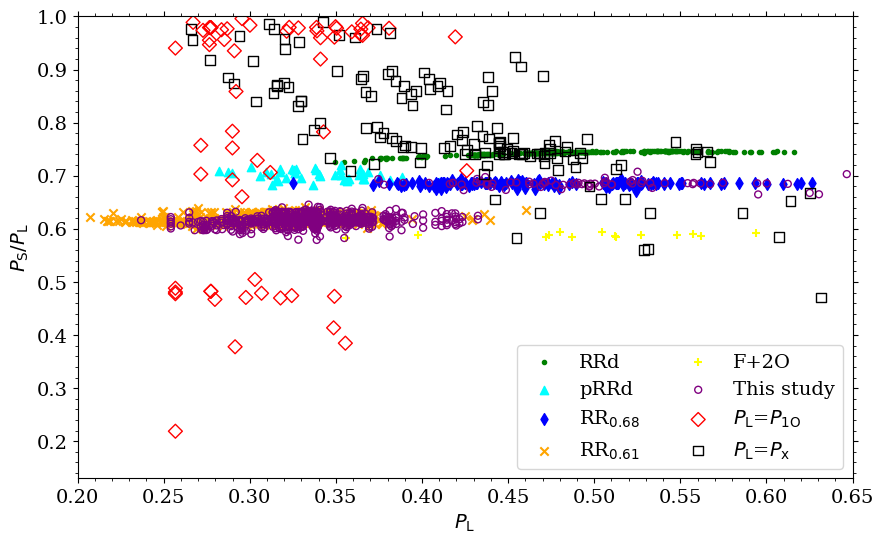}
    \caption{Petersen diagram for additional unknown signals in studied RRc stars. We used red diamonds if the additional detected signal has a period longer than the first-overtone period, and black squared otherwise. We also plotted RR$_{0.61}$ and RR$_{0.68}$ stars identified in this study using purple circles. For comparison we plotted known multi-periodic groups: RRd stars \protect\citep[green points,][]{soszynski2019}, fundamental and second overtone (F+2O) stars \protect\citep[yellow pluses,][]{jurcsik2008,poretti2010,chadid2010,benko2010,benko2014}, peculiar RRd (pRRd) stars \protect\citep[cyan traingles,][]{prudil2017}, and RR$_{0.61}$ and RR$_{0.68}$ from the Galactic bulge \protect\citep[orange crosses and blue diamonds, respectively,][]{netzel_census}.}
    
    \label{fig:additional}
\end{figure}

Besides the well-established groups like RR$_{0.61}$, RR$_{0.68}$ and Blazhko stars, we also found stars with additional signals of mostly unknown origin. We present these stars on the Petersen diagram in Fig.~\ref{fig:additional} together with the already known groups of multi-mode RR Lyrae stars. This includes RRd stars, double-mode RR Lyrae stars pulsating in the fundamental mode and the second overtone; RR$_{0.61}$ and RR$_{0.68}$ stars; and the group of RR Lyrae stars with additional signals of unknown origin discovered by \cite{prudil2017}. We note, that we limited the horizontal axis to 0.65\,d, but there are points extending past that limit. These correspond to stars in which we detected additional signals in a low-frequency range. Such signals are likely due to trends in the data or blending, hence we do not use these as the long-period denominators in the Petersen diagram. They also do not form any specific groups, therefore are not investigated further. 

We also note, that in some stars we detected more than one additional signal of unknown origin. We plotted period ratios where the longer period is the first-overtone with red diamonds. We use black squares to mark stars where the additional signal has a longer period. There are four places in the Petersen diagram where stars seem to cluster. At a period ratio close to 1 there are many stars with additional signals with periods either longer or shorter than the first-overtone period. Such a high period ratio means that the additional signals are located close to the first overtone. The origin of these signals can be due to additional non-radial modes, period changes resulting in close signals or Blazhko modulation that does not fulfill the criteria described in Sec.~\ref{sec.methods} allowing for firm classification as Blazhko.
At period ratios of 0.8--0.9 there is a group of stars in which the longer period belongs to the additional signal. Such signals can be likely explained by very wide subharmonic signals at 0.5$f_{0.61}$ (signals at 0.5$f_{0.61}$ often form complicated and wide structures in frequency spectra, see Fig.~\ref{fig:061example_subharm}). This group in the Petersen diagram consists of 36 stars. Interestingly, 27 stars are also in the RR$_{0.61}$ group, but in the remaining 9 stars, the $f_{0.61}$ signals were not detected. It is possible that in these stars we see the non-radial mode directly. Such stars are already known among RR Lyrae stars \citep{netzel_census,benko2021} and classical Cepheids \citep{rajeev}. 

\begin{figure}
    \centering
    \includegraphics[width=\columnwidth]{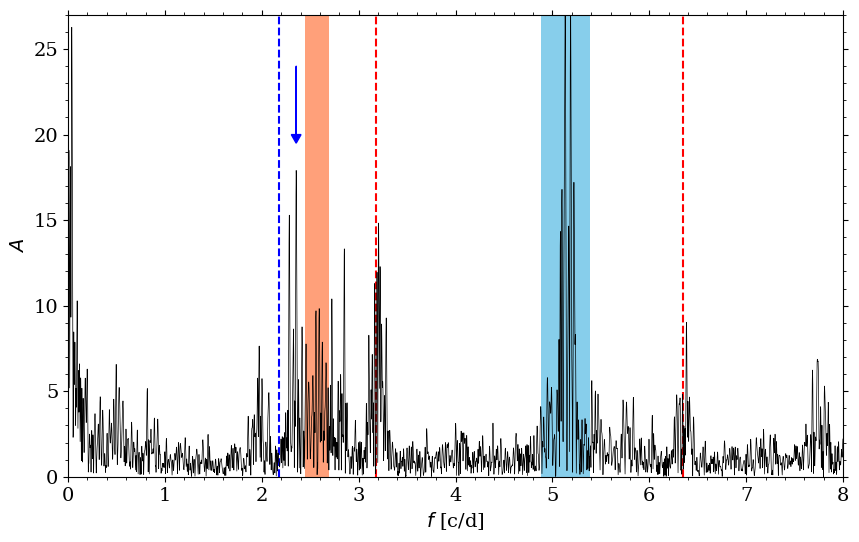}
    \caption{Frequency spectrum for EPIC\,249630037 after prewhitening with the first overtone and its harmonics. Their positions are marked with red dashed lines. The blue dashed line corresponds to the location of $f_{0.68}$ signal (which was not detected in this star). With a blue region, we marked the frequency range of the $f_{0.61}$ signals and with an orange region, we marked the corresponding frequency range of the subharmonic. With a blue arrow, we marked the signal that forms a period ratio that fits the RRd sequence.}
    \label{fig:249630037}
\end{figure}

Equally interesting is a group of stars located above the RRd sequence up to a period ratio of around 0.80 that consists of stars in which the additional signal has longer period than the first-overtone period, too. The origin of such signals is likely to be similar to the 0.80--0.90 group. However, for lower period ratios, when the points get closer to the RRd sequence, it is not clear whether the additional long-period signal is due to a wide subharmonic signal or maybe due to the small-amplitude fundamental mode. In principle, it should be easy to distinguish between the wide structures of subharmonics at 0.5$f_{0.61}$ and coherent signals of the fundamental mode. In Fig.~\ref{fig:249630037} we show that this is not always the case using the frequency spectrum for KIC\,249630037 after prewhitening with the first overtone and its harmonics, whose positions are marked with red dashed lines. The blue region corresponds to the $\langle 0.59,0.65 \rangle P_{\rm 1O}$ frequency range. Corresponding frequency range of the 0.5$f_{0.61}$ signal is plotted in orange. The blue dashed line shows where the (undetected) $f_{0.68}$ would be located. The blue arrow points to the signal that forms a period ratio of 0.7415 with the first-overtone and fits well with the RRd sequence. As it is clearly visible in the frequency spectrum, the signal that would correspond to the fundamental mode is accompanied by many other additional signals that are a part of a group of signals centred at 0.5$f_{0.61}$ signal.

It is also easily visible from Fig.~\ref{fig:249630037} how the two groups at 0.80--0.90-period ratio and between the RRd sequence and the 0.80-period ratio are formed. The former group is created by signals at higher frequencies than the orange region in Fig.~\ref{fig:249630037}. The latter group is formed by signals at frequencies lower than that orange region. It was already predicted and discussed by \cite{dziembowski2016} that signals at subharmonics, i.e., the non-radial mode frequencies, would show significantly more variability and more complex and wider structures than the harmonics. However, it is not clear how wide they can get. This is especially concerning, as the case of the star shown in Fig.~\ref{fig:249630037} illustrates, signals outside the expected frequency range of subharmonics can have higher amplitudes.

\begin{figure}
    \centering
    \includegraphics[width=\columnwidth]{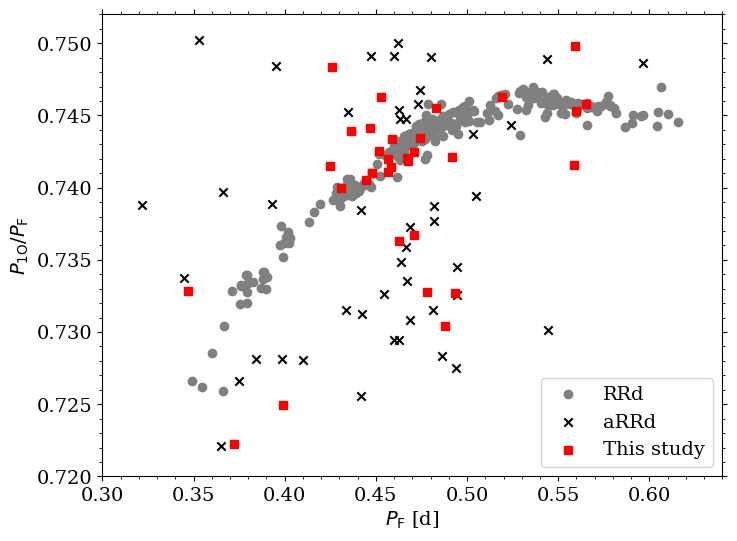}
    \caption{Petersen diagram for RRd stars plotted with grey points and anomalous RRd stars plotted with black crosses \citep{soszynski2019}. With red squares, we plotted stars from this study in which we found additional signals that fit the RRd group in the Petersen diagram.}
    \label{fig:rrd_pet}
\end{figure}

We selected stars that are located at or close to the RRd sequence in Fig.~\ref{fig:additional} as candidates for RRd stars, in which the fundamental mode has a very low amplitude. Selected stars are plotted with red squares in the Petersen diagram in Fig.~\ref{fig:rrd_pet} together with RRd stars and anomalous RRd stars (aRRd)  from the Galactic bulge \citep{soszynski2019}. We selected 32 such stars. 

\begin{figure}
    \centering
    \includegraphics[width=\columnwidth]{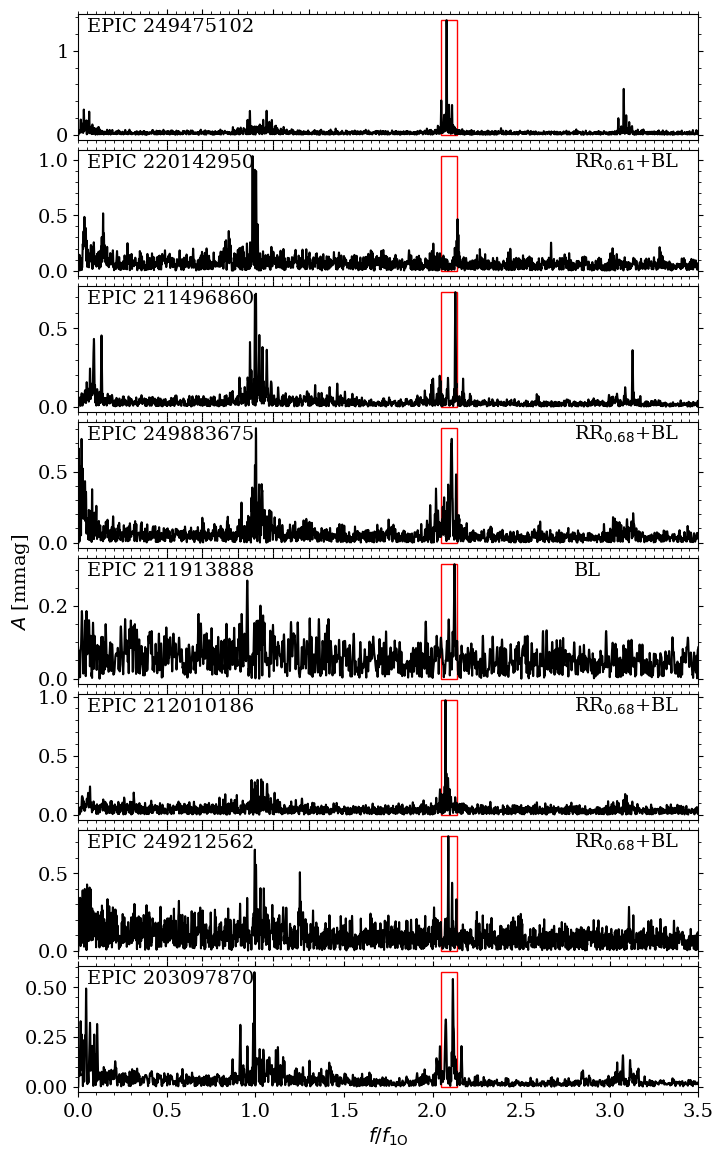}
    \caption{Frequency spectra of eight stars in which we detected additional signals of unknown origin that form a period ratio of around 0.47--0.49 (see text for details). The horizontal axis is scaled with a first-overtone frequency. Red boxes indicate the position of signals forming period ratio in a range of 0.467--0.488. In the upper left corner of each panel, we provided the stars' EPIC number, and in the upper right corner, we provided the classification. If no classification is given, only the unknown signals were detected.}
    \label{fig:047}
\end{figure}

Another group visible in Fig.~\ref{fig:additional} is located at period ratios 0.465--0.490 and consists of 8 stars. The additional signal, $f_{0.48}$, has a period shorter than the first-overtone period and also its first harmonic. Frequency spectra of these eight stars are presented in Fig.~\ref{fig:047} after prewhitening with first-overtone, its harmonics, possible Blazhko sidepeaks, known additional signals such as $f_{0.61}$ or $f_{0.68}$ and associated combination signals or subharmonics. The frequency on the horizontal axis is scaled with the first-overtone frequency. The frequency range of the additional signals, i.e. $\langle 0.467,0.488 \rangle f_{\rm 1O}$, are marked with a red box. In the upper right corner of each panel, we indicated what other signals, if any, were detected in each star. In one star we detected a Blazhko modulation. In three stars we detected the Blazhko effect together with the $f_{0.68}$ signal. In one star we detected Blazhko modulation together with the $f_{0.61}$ signal. In three stars we did not detect anything that would fall into the categories above. The signal $f_{0.48}$ is often accompanied by many signals close to the first-overtone frequency and its harmonics, and sometimes also by signals in the low-frequency range.

\section{Discussion}\label{Sec.discussion}

\subsection{RR$_{0.61}$ stars}
We identified 281 RR$_{0.61}$ stars out of 452 RRc stars, which corresponds to an incidence rate of 62 per cent. This value is higher than incidence rates inferred from analysing ground-based photometry, with exception of the study by \cite{smolec2017} who analysed RR Lyrae stars in the globular cluster NGC~6362 and inferred an incidence rate of 63 per cent. The incidence rate from this study is also slightly lower than values inferred in other studies that used space-based photometry. Analysis of the photometry for the original {\it Kepler} field resulted in the incidence rate of 100 per cent \citep{moskalik2015}. \cite{forro2022} analyzed background RR Lyrae stars in the original {\it Kepler} field and detected four more RRc stars and two new RRd stars. Again, all of them show the 0.61 signal. 
\cite{molnar2015} used K2-E2 data from Pisces and obtained an incidence rate of 75 per cent. \cite{molnar2022} analysed TESS photometric data and found 20 RR$_{0.61}$ stars, which corresponds to the incidence rate of 65 per cent, very similar to that inferred in this study. Different photometric data used in mentioned studies makes the direct comparison of incidence rates not straightforward. We also note that the previous results from space-based photometry are mostly small-number statistics, whereas in this study we used a statistically significant sample.

Even though the incidence rate of RR$_{0.61}$ stars based on this study is slightly lower than inferred from other studies focused on the analysis of space-based photometry, the total number of selected RR$_{0.61}$ stars significantly increases the number of known RR$_{0.61}$ stars observed by space telescopes. Until the time of writing, 30 such stars were known  \citep{gruberbauer2007,chadid2012,szabo2014,moskalik2015,molnar2015,kuehn2017,molnar2022,kurtz2016}. Here we increased this number by an order of magnitude. The analysis of another numerous sample of RR$_{0.61}$ stars observed by the TESS satellite is ongoing (Benk{\H{o}} et al. in prep.).

\begin{figure}
    \centering
    \includegraphics[width=\columnwidth]{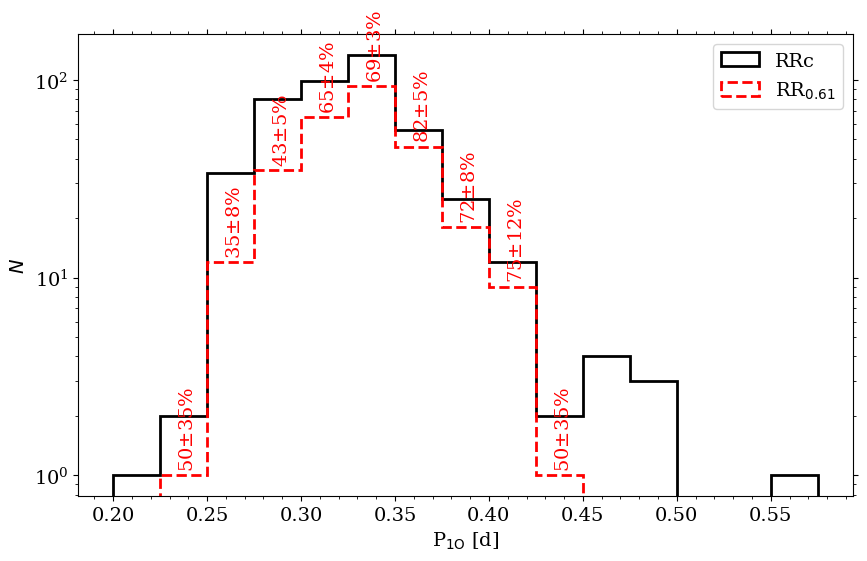}
    \caption{Distribution of first-overtone periods for RRc (black solid line) and RR$_{0.61}$ stars (red dashed line). Incidence rates of RR$_{0.61}$ stars are given for each bin. The errors are calculated assuming the Poisson distribution \citep[e.g.][]{alcock2003}.}
    \label{fig:ir0.61}
\end{figure}

In Fig.~\ref{fig:ir0.61} we plotted the distribution of periods for RRc and RR$_{0.61}$ stars. We also provided the incidence rates of RR$_{0.61}$ for each bin. The errors were calculated assuming Poisson distribution \citep[e.g.][]{alcock2003}. The incidence rate of RR$_{0.61}$ stars is the highest for first-overtone periods between 0.3\,d to 0.4\,d. We did not detect any RR$_{0.61}$ star for first-overtone periods longer than 0.45\,d. We note, however, that even though RRc stars of long periods are known, the long-period stars classified as RRc may be in fact fundamental-mode RR Lyrae stars.

\begin{figure}
    \centering
    \includegraphics[width=\columnwidth]{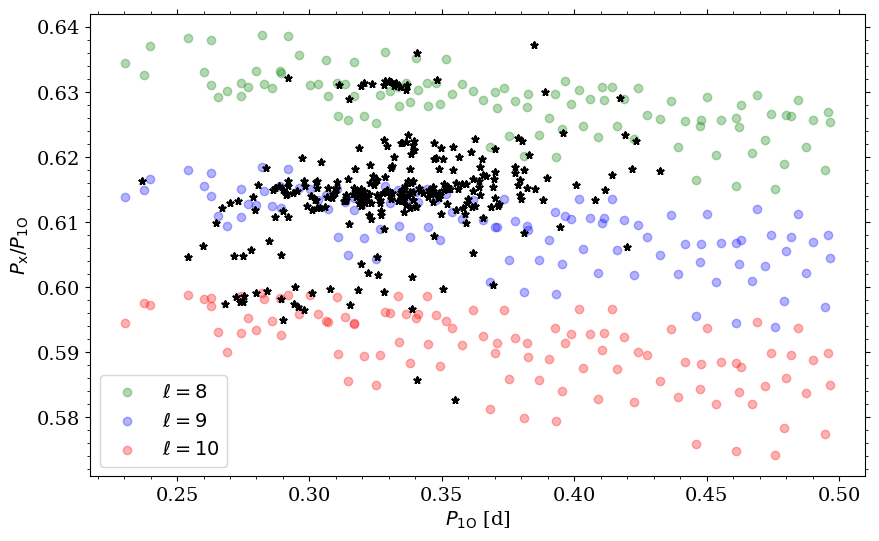}
    \caption{Petersen diagram for RR$_{0.61}$ stars together with theoretical models for RR Lyrae stars calculated with the Warsaw envelope code \protect\citep{dziembowski1977}. Only the highest additional signals are plotted for RR$_{0.61}$ stars, with black symbols. For the modelled period ratios we used the first harmonics of non-radial modes of the given degrees.}
    \label{fig:models}
\end{figure}

In the majority of RR$_{0.61}$ we found more than one additional signal within the period ratio range of 0.57--0.65. The range of period ratios covered by all signals in the RR$_{0.61}$ stars is wide. In particular, many points in the Petersen diagram correspond to a period ratio below 0.61 (see Fig.~\ref{fig:pet061}). Such scatter was observed in stars from the TESS sample analyzed by \cite{molnar2022}, but for a much smaller number of stars. The histogram of period ratios (right panel of Fig.~\ref{fig:pet061}) suggests two additional maxima present below the well-established 0.61 sequence. Particularly interesting is the Petersen diagram when only the highest additional signal in each star is plotted. Two of the well-established sequences at 0.61 and 0.63-period ratios are visible. According to the explanation by \cite{dziembowski2016}, the sequences at 0.63 and 0.61 are due to the harmonics of non-radial modes of degrees $\ell= 8$ and 9, respectively. For the first time, we see another sequence formed at a period ratio of around 0.599. As predicted by \cite{dziembowski2016}, period ratios lower than 0.60 could correspond to the harmonics of non-radial modes of degrees $\ell= 10$ \citep[see fig. 4 in][]{dziembowski2016}. In Fig.~\ref{fig:models} we plotted RR$_{0.61}$ stars using only the highest additional signals and theoretical models for RR Lyrae stars calculated with the Warsaw envelope code \citep{dziembowski1977}. The models were calculated for $M\in(0.5,0.9){\rm M}_\odot$, $\log L/{\rm L}_\odot \in (1.4,1.8)$, $\log T_{\rm eff}\in (3.77,3.9)$, $X \in (0.72,0.75)$ and $Z \in (0.0134,0.000134)$. Only models with a first-overtone period shorter than 0.5\,d and both the first overtone and non-radial modes being linearly unstable are plotted. The models where the $\ell = 10$ non-radial mode is linearly unstable are plotted with red points and cover the range of period ratios corresponding to the lowest sequence at $P_x/P_{\rm 1O}\approx0.599$.

Another possibility that has to be considered regarding the origin of the 0.60 sequence involves a combination signal of the 0.68 mode with the first overtone. An example of a star that has both $f_{0.61}$ and $f_{0.68}$ signals is presented in Fig.~\ref{fig:061_068}. The combination signal, $f_{\rm 1O}+f_{0.68}$, is located close to $f_{0.61}$. However, we consider the explanation of the 0.60 sequence with the combination signals in RR$_{0.68}$ stars unlikely. First, out of the 26 RR$_{0.61}$ stars with a low period ratio, only 5 show also the $f_{0.68}$ signal, too. Second, the RR$_{0.68}$ stars tend to be located close to the period ratio of 0.686 with only a few stars being outliers from the relatively tight 0.686 sequence. Consequently, the sequence created by the combination signal with the first overtone would also show relatively little scatter and would appear tight and horizontal on the Petersen diagram. The average period ratio of the 0.60 sequence is around 0.599, whereas the average period ratio formed by the combination signal of 0.68 signal with the first overtone would be 0.593. 

\begin{figure}
    \centering
    \includegraphics[width=\columnwidth]{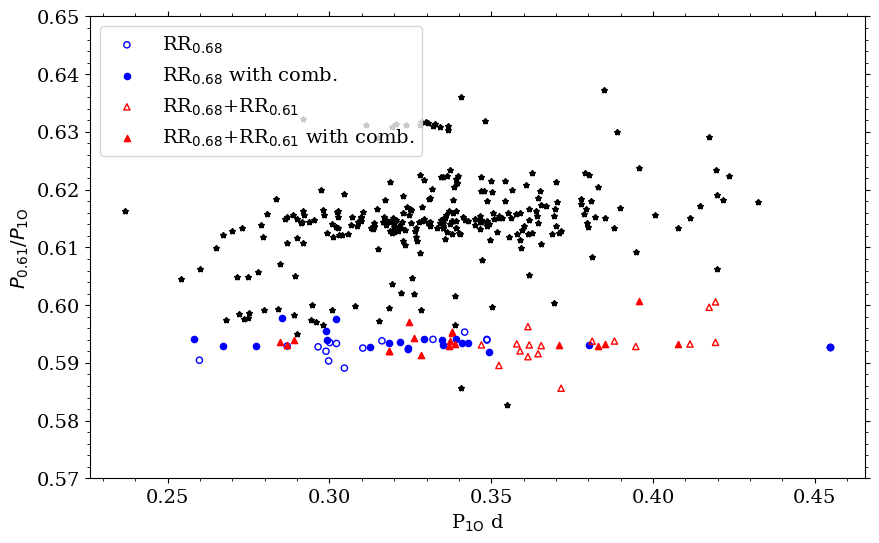}
    \caption{Petersen diagram for RR$_{0.61}$ stars together with RR$_{0.68}$ stars using the combination signal, $f_{\rm 1O}+f_{0.68}$, to calculate the period ratio. Open symbols correspond to RR$_{0.68}$ stars in which we did not detect the combination signal in the frequency spectrum. Filled symbols correspond to stars in which the combination signal was detected. Pure RR$_{0.68}$ stars are plotted with blue circles. Stars that are both RR$_{0.68}$ and RR$_{0.61}$ are plotted with red triangles.}
    \label{fig:comp_0.6sequence}
\end{figure}

This is presented in the Petersen diagram in Fig.~\ref{fig:comp_0.6sequence}. We plotted RR$_{0.68}$ stars using the period ratio that is formed by the combination signal with the first overtone, i.e. $(f_{\rm 1O} + f_{0.68})/f_{\rm 1O}$ with blue and red symbols in the figure. The 0.60 sequence is located indeed close to the sequence created by the combinations frequencies and there is a small overlap. However, the whole sequence cannot be reproduced simply by using the combination signals as there is a significant difference in the average period ratios, indicating that the origin of the $f_{0.68}$ modes is different from that of the $f_{0.61}$ modes.

\subsection{RR$_{0.68}$ stars}\label{sec:disc_0.68}

\begin{figure}
    \centering
    \includegraphics[width=\columnwidth]{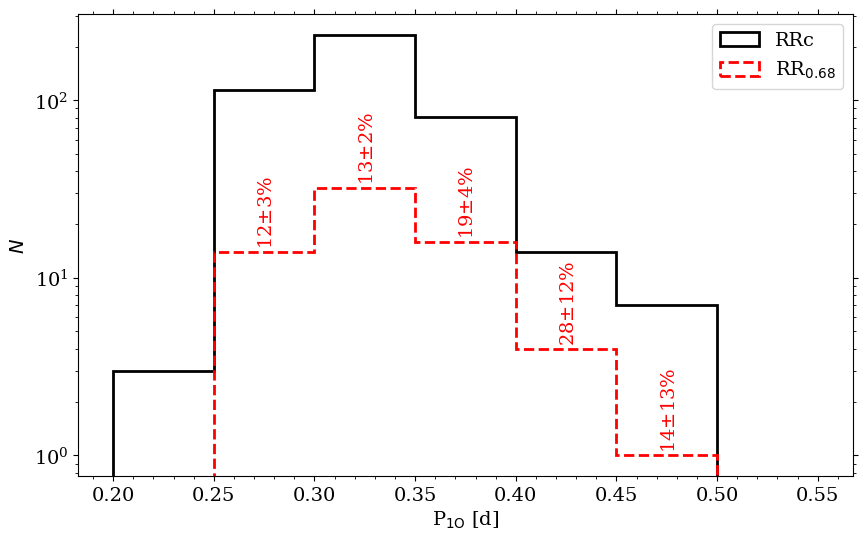}
    \caption{Distribution of first-overtone periods for RRc (black solid line) and RR$_{0.68}$ stars (red dashed line). Incidence rate of RR$_{0.68}$ stars is given for each bin. The errors are calculated assuming the Poisson distribution \citep[e.g.][]{alcock2003}.}
    \label{fig:ir0.68}
\end{figure}

We identified 67 RR$_{0.68}$ stars out of 452 RRc stars, which corresponds to an incidence rate of 15 per cent. This is by an order of magnitude higher than the results from the analysis of ground-based photometry \citep[below 2 per cent,][]{netzel_census}. \cite{molnar2022} analyzed TESS space-based photometry and obtained an incidence rate very similar to this work -- 16 per cent based on five RR$_{0.68}$ stars identified out of 31 RRc stars. The lower incidence rate obtained from the ground-based photometry is not surprising, since the noise level is significantly higher in that data. Additionally, as visible from the period distribution of the RR$_{0.68}$ stars in Fig.~\ref{fig:pet068}, the number of RR$_{0.68}$ stars is likely underestimated for the ground-based data when the period of the additional signal is around 0.5\,d. This corresponds to first-overtone period of around 0.34\,d. In fact, in the K2 sample RR$_{0.68}$ stars that have $P_{0.68} \approx 0.5$\,d are the most numerous. In Fig.~\ref{fig:ir0.68} we plotted the distribution of first-overtone periods for RRc and RR$_{0.68}$ stars together with the incidence rate for each bin. The incidence rate of RR$_{0.68}$ stars increases towards longer first-overtone periods.

The only explanation proposed for RR$_{0.68}$ stars by \cite{dziembowski2016} assumes that these stars are not true RR Lyrae stars, but low-mass (mass of around 0.25\,M$_\odot$) stars stripped from their envelope during the evolution in a binary system similar to the Binary Evolution Pulsator discovered by \cite{pietrzynski2012}. Then the observed period ratio could be explained with double-mode radial pulsations in fundamental mode and first overtone in stripped giants. Stars that show both kinds of the additional signals, $f_{0.61}$ and $f_{0.68}$, challenge this explanation. In this work we significantly increased the number of known stars belonging to both RR$_{0.61}$ and RR$_{0.68}$ groups. The large number of such stars is important for any further attempts to explain the observed signals -- explanation proposed for RR$_{0.68}$ stars has to be consistent with the one proposed for RR$_{0.61}$ stars.

\begin{figure}
    \centering
    \includegraphics[width=\columnwidth]{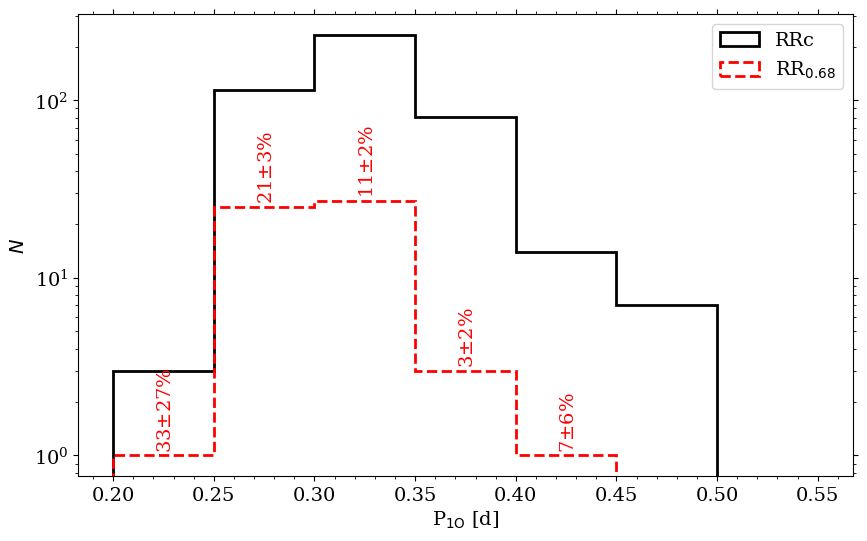}
    \caption{Distribution of first-overtone periods for RRc (black solid line) and Blazhko stars (red dashed line). The incidence rate of Blazhko stars is given for each bin. The errors are calculated assuming the Poisson distribution \citep[e.g.][]{alcock2003}.}
    \label{fig:irbl}
\end{figure}

\subsection{Blazhko stars}

We classified 68 stars as Blazhko stars. In 57 of them the modulation period is covered by the available data. In eleven more stars, the modulation is clearly visible in the light curve, but its period is too long to be fully covered by the available data. The incidence rate is 12.6 per cent when the Blazhko period is covered, or 15 per cent when stars with long modulation periods are included. Both incidence rates are higher than what was derived for RRc stars based on the all but one studies using ground-based photometry. The Blazhko effect in RRc stars was investigated in the Galactic bulge sample by \cite{netzel_blazhko}, \cite{moskalik.poretti2003} and \cite{mizerski2003}. In all of these studies the incidence rate does not exceed 7 per cent. Higher incidence rates of 10 and 19 percent were derived for RRc stars in globular clusters M3 and NGC~6362, respectively \citep{jurcsik2014,smolec2017}. The incidence rate obtained in this study is similar to 13 percent derived by \cite{molnar2022} based on the TESS sample. Lower incidence rates derived using ground-based photometry are easy to explain by the higher noise level in the frequency spectra. Interestingly, the highest observed incidence rate of Blazhko RRc stars is for NGC~6362, even though ground-based data were used for the analysis \citep{smolec2017}. Still, the incidence rate for RRc stars derived in this study is significantly lower than observed for RRab stars \citep[see e.g.][and references therein]{kovacs2016}.

We note, that in many more stars we detected additional signals close to the first-overtone frequency and its harmonics (see Fig.~\ref{fig:additional}). Such signals can be caused by irregular period changes, especially common in RRc stars, and with short data length they could mimic features of the Blazhko modulation. Without longer data we cannot unambiguously classify such stars. There is always a possibility that with longer data those stars could be also classified as Blazhko stars.

In Fig.~\ref{fig:irbl} we plotted the period distribution of RRc stars and Blazhko stars. We also provided the incidence rate of Blazhko stars for each bin. We did not detect Blazhko modulation in stars with a first-overtone period longer than 0.45\,d. The highest incidence rate corresponds to stars with first-overtone periods between 0.25 and 0.30\,d. The incidence rate decreases with an increasing first-overtone period. This observation is in agreement with results from \cite{netzel_blazhko} who also observed the highest incidence rate for the 0.25--0.30\,d first-overtone period.

Four Blazhko RRd stars analyzed by \cite{molnar2022} using TESS photometry painted a complex picture of the Blazhko effect in RRc stars. Low noise level allowed for the detection of low-amplitude sidepeaks. In three out of four stars \cite{molnar2022} observed strong asymmetry in sidepeak amplitudes, i.e., the Q parameter was close to 1.0 or -1.0. Interestingly, in many Blazhko stars in this study, we observed similar asymmetry resulting in a distribution of the Q parameter presented in Fig.~\ref{fig:hist_q}, where many stars have values of the Q parameter close to -1.0. This is in contrast to the results obtained by \cite{netzel_blazhko}, where distribution is asymmetric, centred at $Q\sim -0.1$, and the number of stars decreases with |Q| increasing. In addition, we detected numerous sample of modulated stars with only one sidepeak present, i.e., BL1 stars. Incomplete triplets/multiplets, as in BL1 stars, can be caused by strongly asymmetric sidepeaks or additional non-radial modes with a frequency close to first-overtone frequency. However, we consider the latter scenario unlikely \citep[see discussion in][]{netzel_blazhko}. On the other hand, \cite{benko2011} discussed the scenarios in which the strong asymmetry in amplitudes of sidepeaks can be observed. This is likely the case for the group of stars with $Q\in \langle -1.0, -0.8 \rangle$. We consider BL1 stars to be Blazhko stars with likely strong asymmetry of sidepeak amplitudes. 

Several models have been proposed over the years to explain the mechanism behind the Blazhko modulation. The most promising model predicting modulation of the light curve in RRab stars involves mode resonances, specifically the 9:2 resonance between the fundamental mode and ninth overtone \citep{kollath2021}. 
An important step towards the development of the model behind the Blazhko effect came with the detection of the period doubling in modulated RRab stars \citep{kolenberg2010,szabo2010}. So far, no similar detection was made regarding modulated RRc stars. \cite{netzel_blazhko} tried to detect traces of period doubling using averaged spectra (see their fig.~15). In this study, we attempted a similar approach and also did not detect any signals that might be related to period doubling in Blazhko RRc stars. Thus, the exact mechanism behind the Blazhko effect, in particular in RRc stars, is still an open problem.

\begin{figure}
    \centering
    \includegraphics[width=\columnwidth]{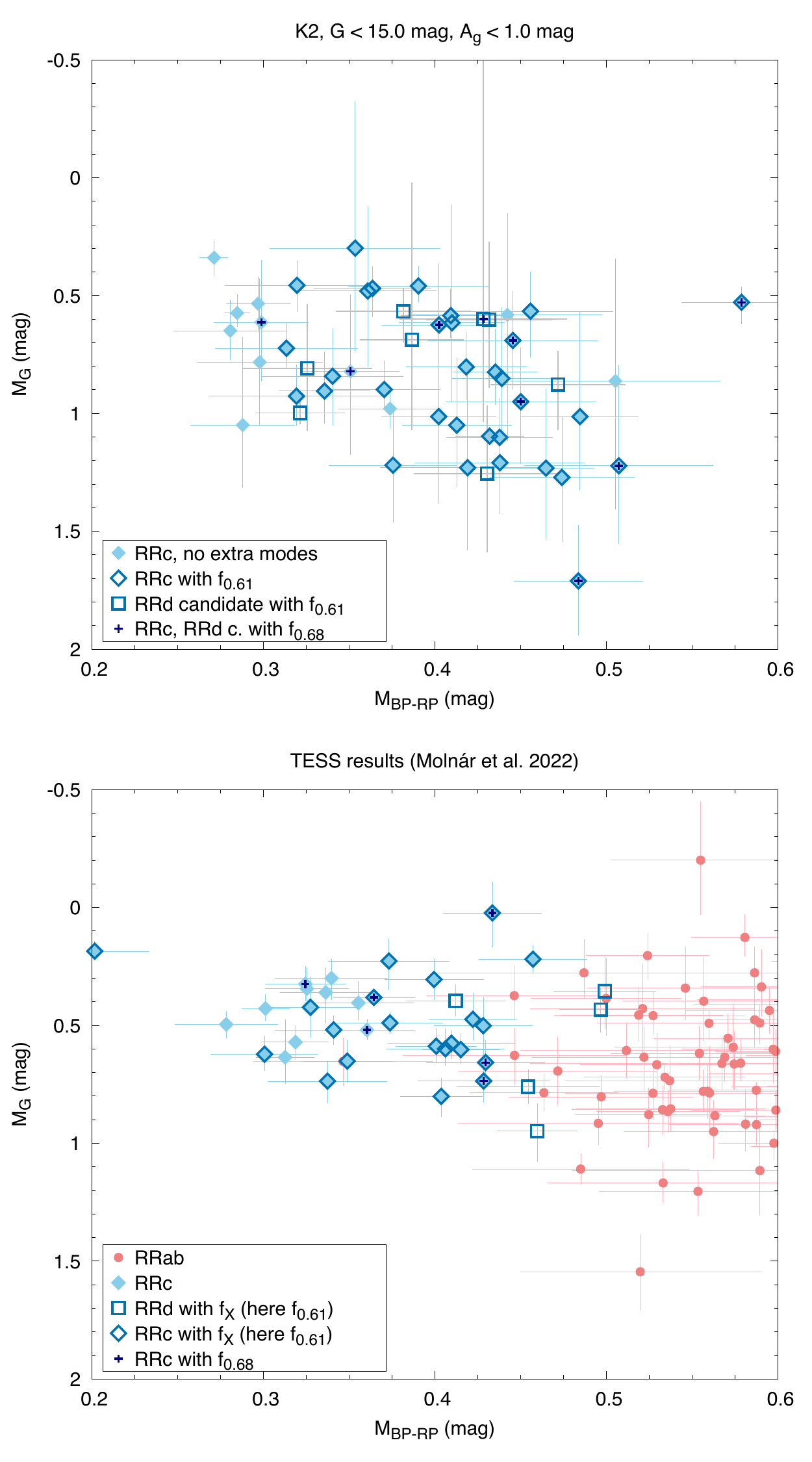}
    \caption{Top: color-magnitude diagram of 53 stars from the bright end of the sample. Presence of additional modes are marked with dark outlines ($f_{0.61}$) and crosses ($f_{0.68}$). RRd candidate stars are marked with squares. Bottom: the CMD of the TESS observations published by \citet{molnar2022}, for comparison. Here, RRab stars are indicated with red symbols.  }
    \label{fig:cmd}
\end{figure}

\subsection{Color-magnitude diagram}\label{sec.cmd}
For a subset of stars we calculated the \textit{Gaia} absolute magnitudes to place them onto an $M_{\mathrm G}$ vs.\ (BP--RP) color-magnitude diagram (CMD). Since the accuracy of \textit{Gaia} parallaxes strongly depends on the apparent brightness of the sources, we limited our sample to stars brighter than 15.0~mag in the \textit{G} band, which translates to a limiting distance of about 7 to 8~kpc \citep{Hernitschek-2019,edr2021}. We used the \texttt{mwdust} package and the \texttt{bayestar} 3D dust map to calculate interstellar extinction in the \textit{Gaia} bands \citep{mwdust-2016}. We then limited the sample to stars where the  extinction in the \textit{G} band ($A_g$) was below 1.0~mag. This left us with 53 stars. 

We marked stars where either the 0.61 or the 0.68 mode(s) are present. The resulting CMD is shown in Fig.~\ref{fig:cmd}, where we plotted the results of \citet{molnar2022} as well for reference. Looking at the two CMDs, we find that the K2 sample contains more stars that are less luminous compared to the TESS sample, extending down to $M_{\rm G} = 1.3$ mag. This range is still in good agreement with the CMD presented by \citet{Clementini-2022}. The difference is likely caused by differences in metallicity: TESS sectors 1 and 2 sampled the halo, whereas some K2 fields also cover the disk that contains metal-rich RR Lyrae stars that are fainter \citep{marconi2015}. 

Both plots indicate that the blue edge of the instability strip is populated by pure overtone pulsators, but cooler RRc stars almost always feature at least one type of extra mode. This is in good agreement with previous observations \citep{jurcsik2015,smolec2017,molnar2022} and with the linear model calculations of \citet{netzel-smolec-2022}. The agreement between the distribution of the observed $f_{0.61}$ modes and the linearly excited high-$\ell$ modes supports the hypothesis of \citet{dziembowski2016} that we are indeed observing high radial-order modes in overtone pulsators.

\begin{figure}
    \centering
    \includegraphics[width=\columnwidth]{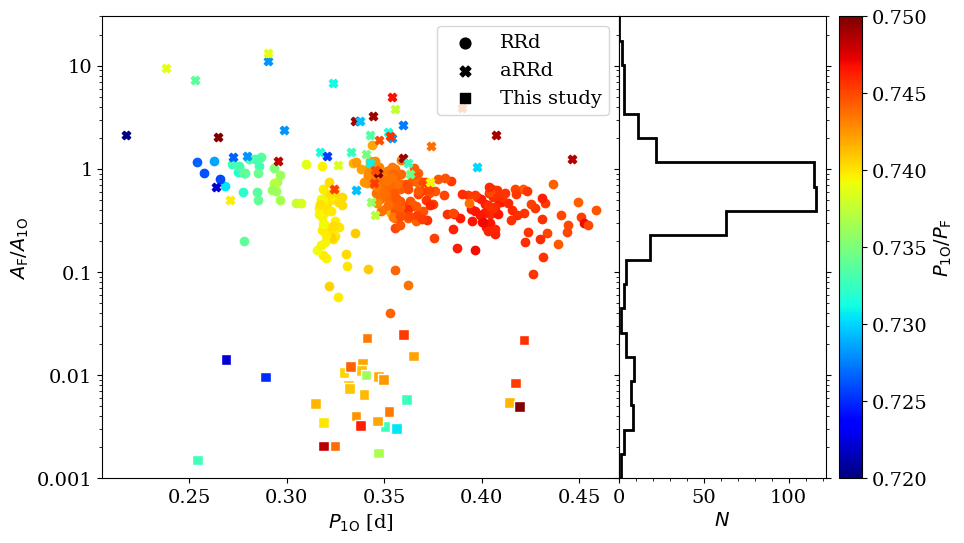}
    \caption{Amplitude ratio of Fourier first-overtone amplitude and the amplitude of the additional signal for RRd candidates (see text for details). The period ratio of the two frequencies is plotted with different colours as indicated in the legend. RRd candidates are plotted with squares. We also included RRd and aRRd stars from the Galactic bulge, which are plotted with circles and crosses, respectively \citep{soszynski2019}. Right panel shows the distribution of the amplitude ratio.}
    \label{fig:rrd_candidates_amp}
\end{figure}

\subsection{RRd candidates}\label{sec:rrd_cand}

We found 32 stars that have additional signals that place them at or close to the RRd sequence (see Fig.~\ref{fig:rrd_pet}). However, in all candidates, the additional signal suspected of being the fundamental mode has a very low amplitude. The amplitude ratio is typically $A_{\rm F}/A_{\rm 1O} < 0.1$, where $A_{\rm 1O}$ is the first-overtone Fourier amplitude and $A_{\rm F}$ is the Fourier amplitude of the additional signal that based on period ratio could correspond to the fundamental mode. In Fig.~\ref{fig:rrd_candidates_amp} we plotted the amplitude ratio, $A_{\rm F}/A_{\rm 1O}$, versus the first-overtone period. Colour corresponds to period ratio on the Petersen diagram. Candidates selected in this study are plotted with squares. For comparison we calculated amplitude ratio for RRd and anomalous RRd stars from the Galactic bulge \cite{soszynski2019}, which are plotted with circles and crosses, respectively. In the right panel of Fig.~\ref{fig:rrd_candidates_amp} we included the distribution of the amplitude ratio. 

Based on the analysis of the TESS sample, \cite{molnar2022} noticed a gap in amplitude ratios (see their fig. 19) between stars with various additional signals and RRd stars. The amplitude ratio presented in Fig.~\ref{fig:rrd_candidates_amp} is bimodal, but the clear gap between RRd stars and RRd candidates is not present.

\begin{figure*}
    \centering
    \includegraphics[width=\textwidth]{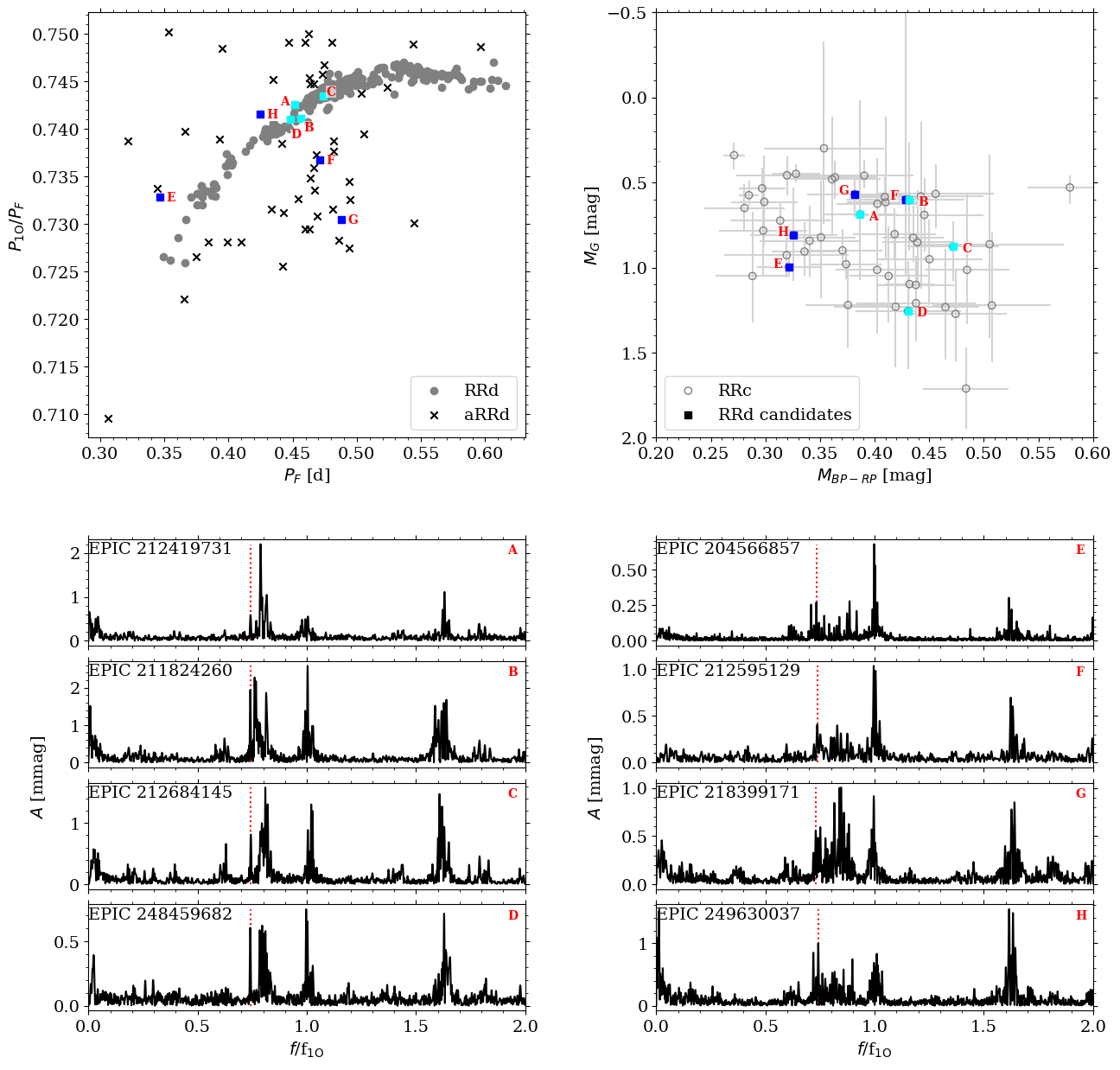}
    \caption{Top left: Petersen diagram showing an RRd sequence (grey points), anomalous RRd stars (black crosses) and RRd candidates that fit the RRd sequence (cyan squares) and that have a slightly atypical period ratio (blue squares). Top right: color-magnitude diagram for RRc stars (black open circles, see also Fig.~\ref{fig:cmd}) and RRd candidates. Bottom left: frequency spectra of four RRd candidates that fit the RRd sequence (cyan squares in the top panels) after prewhitening with the first overtone and its harmonics. The horizontal axis is scaled with the first-overtone frequency. The signal forming period ratio fitting the RRd sequence is marked with a red dashed line. Bottom right: same as the bottom left panel, but for RRd candidates that deviate from the RRd sequence (plotted with blue squares in the top panels). EPIC number is provided in top left corner for each panel with frequency spectrum. Stars on different panels are identified with a letter provided in top right corner of frequency spectra plots and close to the corresponding points in the two upper plots.} 
    \label{fig:rrd_bright}
\end{figure*}

To verify whether these stars might be RRd stars, we plotted them on the CMD. Again, as in Sec.~\ref{sec.cmd}, we limited the sample to stars brighter than 15.0 mag in the $G$ band, which leaves us with 8 RRd candidates. The 8 stars are plotted in the Petersen diagram and CMD in Fig.~\ref{fig:rrd_bright} with square symbols. Four stars (cyan squares) fit very well to the progression of the RRd sequence in the Petersen diagram. The other four stars (blue squares) are outliers from the RRd sequence. Interestingly, the distinction between these two groups is connected with their positions on the CMD. Stars that follow the RRd sequence are redder. Stars that are outliers from the sequence are typically bluer, likely too blue to classify them as RRd stars. 

In the lower panels of Fig.~\ref{fig:rrd_bright} we plotted frequency spectra of the 8 RRd candidates. In the left column, we plotted stars that follow the RRd sequence (cyan squares in the Petersen diagram) and in the right column, we plotted stars that are outliers (blue squares). Frequency spectra were prewhitened with the first overtone and its harmonics as well as any other additional signals with exception of $0.5f_{0.61}$ subharmonics and signals in its vicinity. We note that in all eight stars the signals at $f_{0.61}$ were detected together with the subharmonics. We marked the signal that based on the period ratio might correspond to the fundamental mode with red dashed lines. In all stars, the additional signal is located close to the subharmonic at $0.5f_{0.61}$. Especially in the case of stars that are outliers from the RRd sequence, the additional signal blends with the wide structure of the subharmonic. Interestingly, the separation between the signal in question and the subharmonic is more noticeable in the frequency spectra of stars from the cyan group.

Of all 32 stars that are RRd candidates, 16 follow the RRd sequence (see Fig.~\ref{fig:rrd_pet}). However, we cannot check their position on the CMD to verify their classification. 

One of the reasons to doubt the nature of the additional signal is the fact that in the majority of RRd candidates we also observe the $f_{0.61}$ signal together with its subharmonics. Indeed, only in two stars from 32 RRd candidates did we not detect the $f_{0.61}$ signal. In one of the two stars, EPIC\,246159987, we detected two signals. The first one, $f_1=2.17862522$\,d, forms a period ratio of  0.743 with the first-overtone and fits the RRd progression. The second signal $f_2=2.41719765$\,d forms period ratio around 0.825. Interestingly, the (not observed) harmonic of $f_2$ would form a period ratio around 0.606 with the first overtone, which means that the $f_2$ signal might be a detection of non-radial mode corresponding to RR$_{0.61}$ stars. In the second star, EPIC\,249605297, we detected Blazhko modulation and additional signal, $f_1=2.68763533$ d, which forms a period ratio of around 0.722 which does not fit the RRd sequence. However, the period ratio is consistent with anomalous RRd stars, that show atypical period ratios and modulation \citep{soszynski.smolec2016}. The signal $f_1$ forms combination signals with the first overtone. The amplitude ratio is low, $A_1/A_{\rm 1O} \approx 0.014$, which is the opposite to aRRd stars. Namely, in aRRd stars the fundamental mode typically has a higher amplitude than the first overtone. Similar RRd candidates were already reported by \cite{molnar2022}. The final classification of these stars as RRd or aRRd stars remains uncertain.


\section{Conclusions}\label{Sec.conclusions}

We analyzed 3917 light curves of 3057 objects from K2 Campaigns 0--19. We classified 452 stars as RRc stars and performed frequency analysis on them to search for additional phenomena such as the presence of non-radial modes and/or modulation.

The most important results we obtained from these high-accuracy, continuous light curves are the following:
\begin{enumerate}
    \item We detected 281 RR$_{0.61}$ stars. This not only increases the total number of known RR$_{0.61}$ stars but most importantly, significantly increases the number of such stars observed from space. The incidence rate of RR$_{0.61}$ stars in our sample is 62 per cent.
    \item Besides the well-known three sequences formed by the RR$_{0.61}$ stars in the Petersen diagram, we found indications of a fourth sequence at the period ratio of around 0.59. A possible explanation of this sequence would be as harmonics of non-radial modes of degree $\ell=10$.
    \item We detected 67 RR$_{0.68}$ stars, which constitute the largest sample of these stars observed by the space telescope. The incidence rate of RR$_{0.68}$ stars in our sample is 15 per cent. 
    \item In 32 stars we found signals corresponding to both RR$_{0.61}$ and RR$_{0.68}$ groups. Such stars are very important from the perspective of explaining the nature of RR$_{0.61}$ and RR$_{0.68}$ stars simultaneously.
    \item We found Blazhko modulation in 57 stars, where the modulation period is within the data length. In 11 more stars, the amplitude modulation is visible in the light curve, but the period is too long to be covered by the data. The incidence rate is 12.6\% in the former case and 15\% in the latter.
    \item We found 32 stars with additional signals that put these stars on or in the vicinity of the RRd sequence in the Petersen diagram. These RRd candidates would feature the fundamental mode with a very low amplitude. However, the classification of them as RRd stars remains uncertain.
    \item We found a new group formed by eight stars in the Petersen diagram around the period ratio of 0.465--0.490. The origin of those signals is unknown.
    \item  In 125 stars we detected signals that do not fall into the above mentioned groups or do not correspond to known instrumental signals. For the majority of them, the origin of the additional signals remains unknown.
\end{enumerate}

This study represents the most detailed investigation of the mode content of RRc stars via space-based photometry so far. Together with the numerical advancements of the field, we are now able to make asteroseismic inferences for RR Lyrae stars and we getting closer to understand the additional modes observed in them.   

\begin{acknowledgements}
H.N has been supported by the \'UNKP-22-4 New National Excellence Program of the Ministry for Culture and Innovation from the source of the National Research, Development and Innovation Fund. 
This project has been supported by the Lend\"ulet Program of the Hungarian Academy of Sciences, project No. LP2018-7/2020, by the `SeismoLab' KKP-137523 \'Elvonal and  NN-129075 grants of the Hungarian Research, Development and Innovation Office (NKFIH). This work uses frequency analysis software written by R. Smolec. This paper includes data collected by the K2 mission. Funding for the \textit{Kepler} and K2 missions is provided by the NASA Science Mission Directorate. This work has made use of data from the European Space Agency (ESA) mission {\it Gaia}, processed by the {\it Gaia} Data Processing and Analysis Consortium (DPAC). Funding for the DPAC has been provided by national institutions, in particular the institutions participating in the {\it Gaia} Multilateral Agreement. This research has made use of NASA's Astrophysics Data System (ADS).

\end{acknowledgements}

\bibliographystyle{aa} 
\bibliography{references} 

\appendix
\section{Dubious classification}\label{Sec.dubious}
In several stars from our sample, the simple classification as described in Sec.~\ref{sec.methods} was unambiguous. Those stars are described below:

\paragraph{EPIC\,214602901}
In this star we found an additional signal that fits the RR$_{0.61}$ group. After prewhitening with this signal and its subharmonic, in frequency spectrum remains strong signal unresolved with the first overtone. Visual inspection of light curve shows decrease in the amplitude, which is not as clear as in the Blazhko stars. We checked temporal variations in amplitude and phase of the first overtone. They clearly shows variability and decrease in the amplitude, which are, however, not as periodic as one would expect from the Blazhko stars. Classification of this star as Blazhko star remains uncertain.

\paragraph{EPIC\,251562169}
This star has three close additional signals forming period ratio 0.60 -- 0.61 and is classified as RR$_{0.61}$. However, those signals do not have their corresponding signals at subharmonic frequencies. Close to the frequency range of subharmonic at half of the frequency we found signals but have slightly higher or lower frequencies than expected. Three signals located there form period ratios of around 0.76, 0.87 and 0.89, therefore they do not fit well into any other known multi-mode group. 

In this star we also detected a few additional signals close to the first-overtone frequency. They do not form clear multiplets connected to the Blazhko effect. Visual inspection of the light curve does not indicate any kind of modulation. Time variation of amplitude and phase of the first overtone are neither periodic enough nor correlated, therefore this star was not classified as a Blazhko star.

\paragraph{EPIC\,203097870}
In this star we detected signals close to the first-overtone frequency and its harmonics, but they do not form clear multiplets. We also detected strong signals in the low frequency range. Light curve show no clear modulation. Modulation is however visible in time variation of amplitude and phase of the first overtone. We note that this star is a likely Blazhko candidate, but due to the complicated frequency spectrum we did not classify this star as Blazhko yet.

\paragraph{EPIC\,214145992}
In this star the highest additional signal, $f_{0.68}=2.31504082$ d$^{-1}$ corresponds to the RR$_{0.68}$ group. In frequency spectrum we found three more signals. First one, at frequency $f_1=4.35728498$ d$^{-1}$, would fit the combination frequency between $f_{0.68}$ and first overtone, $f_{\rm 1O}$, but in a form $2f_{\rm 1O}-f_{0.68}$, whereas the simple combination $f_{\rm 1O}+f_{0.68}$ is not present in the frequency spectrum. Signal at $f_1=4.35728498$ d$^{-1}$ is also unlikely explained by the instrumental signal expected at $f \approx 4.07$ d$^{-1}$. Interestingly, $f_{0.68}$ and $f_1$ are almost equally spaced with respect to $f_{\rm 1O}$ with the separation of $\Delta f \approx 1.02$ d$^{-1}$. Two more additional signals are found at $f_2=7.41474260$ d$^{-1}$ and $f_3=0.74308575$. They are likely aliases of the first-overtone of separation $\Delta f \approx 4.07$ d$^{-1}$.

\paragraph{EPIC\,248765005}
This star is classified as RR$_{0.68}$ due to the additional signal at $f_{0.68}=2.16303756$ d$^{-1}$. Similarly to EPIC\,214145992, we found another signal, $f_1$, that would fit the combination with the first-overtone frequency, $f_{\rm 1O}$, in the form $2f_{\rm 1O}-f_{0.68}$. Combination frequency in the form $f_{\rm 1O}+f_{0.68}$ was not detected. Again, as in the case of EPIC\,214145992, the separation between the first-overtone and $f_{0.68}$, and the first-overtone and $f_1$ are almost equal and $\Delta f \approx 0.999$ d$^{-1}$. The signal at $f_1 = 4.16167610$ d$^{-1}$ is close to the position of the possible instrumental signal, however typically the instrumental signal is located at frequency $\approx 4.07$ d$^{-1}$.

\paragraph{EPIC\,249896645} This star has a very rich frequency spectrum.  We detected signal corresponding to the RR$_{0.68}$ group, forming period ratio with the first overtone around $0.684$. In the case of the $f_{0.68}$ signal we faced the same problem as described in the case of EPIC\,214145992 and EPIC\,248765005. In the case of this star, the signal $f_1$ forming the puzzling combination $2f_{\rm 1O}-f_{0.68}$ is located at 3.45198218 d$^{-1}$, therefore cannot be explained with the instrumental signal. The separation between the first-overtone frequency and $f_{0.68}$ or $f_1$ is $\approx 0.829$ d$^{-1}$. 

We also found signals corresponding to RR$_{0.61}$ group. Four signals are detected in the interesting frequency range. They form period ratios of 0.608, 0.621, 0.634 and 0.585. Interestingly, only the signal corresponding to the lowest period ratio has a corresponding subharmonic at half of its frequency. 

Close to the frequency range of subharmonics, we detected another signal, that forms period ratio with the first overtone of around 0.768, so it does not fit any particular group in the Petersen diagram. Another detected additional signal is located at $f=1.47$ d$^{-1}$ and forms period ratio of around 0.56, which also does not correspond to any known multi-mode group. 

We also detected many signals close to the first-overtone frequency, $f_{\rm 1O}=2.62337858$ d$^{-1}$. Namely, signals at $f_a=2.63698090$ d$^{-1}$, $f_b=2.74156059$ d$^{-1}$, $f_c=2.68927643$ d$^{-1}$, $f_d=2.40492376$ d$^{-1}$ and $f_e=2.84183339$ d$^{-1}$. They do not form simple multiplets that can be immediately connected with the Blazhko modulation. However, visual inspection of the light curve shows slight amplitude change. 

\paragraph{EPIC\,251248707}
In this star we detected the Blazhko sidepeaks and two independent additional signals. The additional signal of the highest amplitude, $f_1$, forms period ratio with the first-overtone period of 0.597. We also detect the signal that can be interpreted as linear combination in the form $f_1+f_{\rm 1O}$. The second additional signal, $f_2$, forms period ratio with the first-overtone around 0.685. However, it could be also explained as linear combination in the form $f_1-f_{\rm 1O}$. The frequency of such linear combination is unresolved with the detected signal, $f_2$. The frequency difference between $f_2$ and calculated frequency of the linear combination corresponds to 30\,d, whereas the data length is 47\,d. We classified this star as RR$_{0.61}$.

\paragraph{EPIC\,201552850}
This star is classified as a RR$_{0.68}$ star. It also shows clear modulation in the light curve. Modulation is likely multiperiodic, with the longest period too long for the used data. We included this star in the Blazhko group as well (Table~\ref{tab:bl}), but only using the period of modulation that is covered by the data. We note, that the multi-periodic modulation visible in the light curve results in the complex structure in frequency spectrum involving sidepeaks unresolved with the first-overtone and therefore not prewhitened during the analysis.

\paragraph{EPIC\,210740526}
This star is member of both RR$_{0.61}$ and RR$_{0.68}$ groups. It also shows many signals close to the first-overtone frequency, which do not form simple multiplets. Time variations of amplitude and phase of the first overtone are present, however they are not periodic enough to indicate the Blazhko effect. Longer data coverage would be necessary to draw any conclusions.

\end{document}